\documentclass[%
 aip,
 amsmath,amssymb,
 reprint,%
]{revtex4-1}

\usepackage{graphicx}
\usepackage{dcolumn}
\usepackage{bm}

\usepackage[utf8]{inputenc}
\usepackage[T1]{fontenc}
\usepackage{mathptmx}
\usepackage{etoolbox}

\makeatletter
\def\@email#1#2{%
 \endgroup
 \patchcmd{\titleblock@produce}
  {\frontmatter@RRAPformat}
  {\frontmatter@RRAPformat{\produce@RRAP{*#1\href{mailto:#2}{#2}}}\frontmatter@RRAPformat}
  {}{}
}%
\makeatother

\preprint{AIP/123-QED}

\begin{document}


\title{Predicting Electromagnetically Induced Transparency based Cold Atomic Engines using Deep Learning}

\author{Manash Jyoti Sarmah}
 \altaffiliation[Also at ]{Department of Chemistry, Gauhati University, Gopinath Bordoloi Nagar, Jalukbari, Assam, India.}
\author{Himangshu Prabal Goswami}%
 \email{hpg@gauhati.ac.in}
\affiliation{%
Department of Chemistry, Gauhati University, Gopinath Bordoloi Nagar, Jalukbari, Assam, India.\\
}%

\date{\today}




\begin{abstract}
We develop an artificial neural network model to predict quantum heat engines working within the experimentally realized framework of electromagnetically induced transparency. We specifically focus on \( \Lambda \)-type alkali-based cold atomic systems. This  network allows us to analyze all the alkali atom-based engines' performance.  High performance engines are predicted and analyzed based on three figures of merit: output radiation temperature, work and ergotropy. Contrary to traditional notion, the algorithm reveal the limitations of output radiation temperature as a stand alone metric for enhanced engine performance.  In high-output temperature regime, Cs-based engine with a higher output-temperature than Rb-based engine is characterized by lower work and ergotropy. This is found to be true for different atomic engines with common predicted states in both high and low-output temperature regimes. Additionally, the ergotropy is found to exhibit a saturating exponential dependency on the control Rabi frequency. 
\end{abstract}


\maketitle


\section{\label{sec:level1}Introduction}
The foundational theoretical models of quantum heat engines (QHEs), exemplified by the pioneering three-level quantum system proposed by Scovil and Dubois~\cite{scovil1959three_qt_th, PhysRevLett.2.262_scovil_1st}, heralded a transformative era in quantum thermodynamics—an interdisciplinary domain bridging quantum mechanics with classical thermodynamics. This model demonstrated that the efficiency of a QHE, much like its classical counterparts, is fundamentally constrained by the Carnot limit~\cite{PhysRevA.88.013842_hpg01} and established the quantum analog of classical heat engines. QHEs are typically modeled using two-~\cite{PhysRevB.110.134318_2_lev_2024}, three-~\cite{deng2024capturingdynamicsthermodynamicsthreelevel_3_lev_2024}, or more-level quantum systems, spin systems~\cite{https://doi.org/10.1002/andp.202400143_mriga}, or harmonic oscillators~\cite{10.1116/5.0072067_HO_2022} as the working medium. The working medium is coupled to thermal reservoirs at different or same temperatures, allowing energy exchange that drives a thermodynamic cycle, akin to classical heat engines. Recent studies, including advancements in understanding efficiency fluctuations~\cite{PhysRevA.110.052214_mjs_efficiency, PhysRevA.107.L031302_fluctuations_exp}, noise-induced coherences~\cite{ PhysRevA.88.013842_hpg01, SARMAH2023129135}, and ergotropy~\cite{PhysRevA.110.032213_mjs_ergotropy, van2020single_exp_erg} have underscored the role of QHEs as promising candidates for next-generation energy technologies. 


Building on this foundation, modern research has shifted towards exploiting advanced quantum phenomena to enhance the performance and versatility of QHEs. A particularly promising direction involves the integration of electromagnetically induced transparency (EIT) into QHE design. EIT is a quantum optical phenomenon where destructive interference between atomic excitation pathways suppresses absorption at certain frequencies, rendering the working medium transparent to specific photons~\cite{boller1991observation, harris1990nonlinear} which is a direct consequence of quantum coherences. EIT-based QHEs leverage on this coherence to precisely control light-matter interactions, allowing for increased efficiency in photon exchange~\cite{PhysRevA.94.053859_seharris, PhysRevA.109.012207_Zhang}. Fine-tuning the coherence in an EIT system enables the medium to operate beyond the constraints of Kirchhoff's law~\cite{PhysRevA.94.053859}. EIT-based QHEs, particularly in single-atom QHEs operated within optical cavities, have demonstrated coherent coupling mechanisms to enhance photon generation and emission processes~\cite{PhysRevA.109.012207_Zhang}. Composite QHEs, which integrate nanomechanical mirrors and ultracold atomic gases, have also demonstrated increased brightness under EIT conditions due to mirror-induced modulations in output radiation~\cite{Laskar_2024}. Additionally, gain-assisted QHEs utilizing spontaneously generated coherence have shown improved emission cross-sections and output brightness, showcasing the versatility of EIT in optimizing QHE~\cite{NIAZ2024171816}. A landmark experimental demonstration of EIT-based QHEs~\cite{PhysRevLett.119.050602_zou} used a three-level $\Lambda-$ configuration of cold rubidium ($^{85}$Rb) atoms confined in a two-dimensional magneto-optical trap (MOT). Thermal reservoirs were simulated by modulating laser beams with white noise, which introduced random phase components, effectively reproducing thermal environments. Under EIT conditions, this setup allows the atomic medium to become transparent to a probe frequency at the line center and measurements of photon emissions revealed a brightness nearly nine times greater than that of the ambient reservoir temperature, confirming theoretical predictions on the enhanced performance of EIT-based QHEs~\cite{PhysRevA.94.053859_seharris}. These advancements underscore the versatility of EIT in optimizing QHE performance, particularly through improvements in emission brightness and quantum coherence control. One can optimize the performance of EIT-based QHEs by exploring their parameter space, i.e., by carefully selecting atomic states and transitions that produce more output brightness. Such optimizations could enable more efficient energy extraction, underscoring the need for a comprehensive analysis of atomic transitions and configurations. However, despite these advances, the development of practical EIT-based QHEs remains challenging due to the extensive parameter space that must be explored to optimize atomic transitions and configurations. Leveraging on machine learning approaches could help identify optimal atomic configurations of coherence, interaction strengths, and environmental conditions. In this context, artificial neural networks (ANNs) offer a powerful tool for efficiently mapping the relationships between atomic parameters and EIT characteristics, facilitating the effective screening of potential atomic configurations that maximize QHE performance.

In this work, ANNs are employed to analyze datasets encompassing atomic parameters and EIT properties across various alkali metals. By training the ANNs on labeled data, we enable accurate predictions of EIT-based QHEs that exhibit enhanced output performance. This approach reduces the computational burden of parameter exploration and accelerates the identification of configurations with high potential for practical applications in micro-scale power generation and cooling.
The structure of this paper is as follows: Section~\ref{sec:Model} introduces the general model of a $\Lambda$-type QHE employing EIT. Section~\ref{nn_mapp} outlines the neural network (NN) mapping technique. Section~\ref{Modelling and Data Generation} accounts for the data generation process and artificial neural network (ANN) modeling. Section~\ref{ANNresults} discusses the performance of the ANN models. Section~\ref{predict} highlights the predictions made by the models and explores their application in identifying optimal atomic systems for EIT-based QHEs. Finally, conclusions are drawn in Section~\ref{conclusion}.

\section{\label{sec:Model}The Model}

\begin{figure}[h!]
    \centering
    \includegraphics[width=0.4\textwidth]{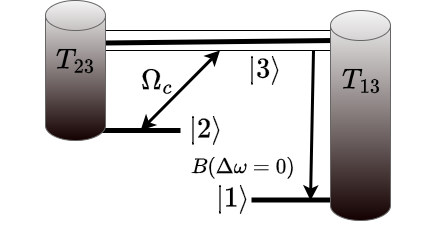}
    \caption{Energy level diagram for a three-level \( \Lambda \)-type QHE system with states \( |1\rangle \), \( |2\rangle \), and \( |3\rangle \). States \( |1\rangle \) and \( |3\rangle \) are in contact with a reservoir at temperature \(T_{13}\) and States \( |2\rangle \) and \( |3\rangle \) are in contact with a reservoir at temperature \(T_{23}\). The double-headed arrow indicates the Rabi frequency (\( \Omega_C \)) between states \( |2\rangle \) and \( |3\rangle \), while the single-headed arrow denotes output spectral brightness, \(B(\Delta\omega)\), from the highest energy state \( |3\rangle \) to the ground state \( |1\rangle \).}
    \label{fig:QHE_levels}
\end{figure}

The atomic configuration analyzed in this study, illustrated in Fig.~\ref{fig:QHE_levels}, is a three-level \( \Lambda \)-type system. Such systems have been extensively investigated as paradigms for generating coherent radiation without population inversion~\cite{PhysRevLett.64.1107_Imamo_2, RevModPhys.77.633_Imamogl_1, PhysRevLett.66.1154_Imamo_3, scovil1959three_qt_th, PhysRev.156.343_Geusic, PhysRevLett.119.050602_zou, harris1990nonlinear, PhysRevA.94.053859}. The system operates as a closed quantum configuration, where the dynamics are primarily governed by incoherent pumping and spontaneous decay processes. The \( |1\rangle \to |2\rangle \) transition is metastable with a transition frequency \( \omega_{12} \). A monochromatic ``coupling" laser with Rabi frequency \( \Omega_C \) is applied at the line center of the \( \omega_{23} \) transition, introducing coherence to the system. Blackbody radiation at a temperature \( T_{13} \) interacts with the \( |1\rangle \to |3\rangle \) transition, while radiation at a temperature \( T_{23} \) interacts with the \( |2\rangle \to |3\rangle \) transition. These temperatures can differ or be identical, with either being higher than the other. 
The quantum heat cycle can be described as \( |1\rangle \xrightarrow{T_{13}} |3\rangle \xrightarrow{T_{23}} |2\rangle \xrightarrow{\Omega_c} |3\rangle \xrightarrow{\omega} |1\rangle \). The overall process is as follows: Initially, a photon is absorbed from the \( T_{13} \) reservoir, causing an electron in the ground state \( |1\rangle \) to excite to the upper state \( |3\rangle \). As a result, a photon is emitted into the \( T_{23} \) reservoir, increasing the population of state \( |2\rangle \) by one unit. Subsequently, a photon is absorbed from the coupling laser, exciting the atom from the metastable state \( |2\rangle \) to the upper level \( |3\rangle \). In the final step, a photon of energy \( \omega \) is emitted during the \( |3\rangle \to |1\rangle \) transition, representing the output of the QHE. This interplay between incoherent and coherent interactions forms the foundation for the enhanced performance of the QHE~\cite{PhysRevLett.119.050602_zou, PhysRevA.94.053859_seharris}. The Hamiltonian of the system is provided in the Supplementary Information 1. We assume that the field on the \(|1\rangle \rightarrow |3\rangle\) transition is sufficiently weak such that the populations are determined by the driving rates \(R_{ij}\) and the strong coupling field \(\Omega_C\) and obtain \(\dot{\rho}_{ii}\) as follows~\cite{PhysRevLett.119.050602_zou, RevModPhys.77.633_Imamogl_1, PhysRevA.94.053859_seharris, NIAZ2024171816}:
\begin{align*}
R_{13} \rho_{33} - R_{13} \rho_{11} &= 0, \\
R_{23} \rho_{33} -( R_{23} + \Omega_C ) \rho_{22}  &= 0, \\
R_{13} \rho_{11} + (R_{23} + \Omega_C) \rho_{22} - (R_{23} + R_{13}) \rho_{33} &= 0, \\
\rho_{11} + \rho_{22} + \rho_{33} &= 1,
\end{align*}
where \( R_{13} \) and \( R_{23} \) are the transition rates. The rates \( R_{ij} = R_{ji} \) are related to the thermal occupation numbers \(\bar{n}_{ij}\) by \( R_{ij} = \Gamma_{ij} \bar{n}_{ij} \), where \(\omega_{ij}\) are the transition frequencies and \(\Gamma_{ij}\) are the lifetime decay rates with \(i,j \in (|1\rangle, |2\rangle, |3\rangle)\).The output of the QHE is characterized by the spectral brightness \( B(\omega, z) \), which represents the number of photons per second generated in the \( z \)-direction for a single transverse mode~\cite{PhysRevA.94.053859_seharris, PhysRevLett.119.050602_zou, PhysRevLett.66.1154_Imamo_3}. The maximum brightness is given by, \( B(\omega) \),

\begin{align}
    B(\omega) = \frac{\Theta\sigma_{\text{em}}}{\sigma_{\text{abs}} - \Theta\sigma_{\text{em}}},
\end{align}
where \(\sigma_{abs}\) and \(\sigma_{em}\) are the absorption and emission cross sections~\cite{0-19-503437-6_b_w_deri_book, PhysRevA.94.053859_seharris, PhysRevLett.66.1154_Imamo_3} and \(\Theta\) is defined as the ratio of atoms in the upper manifold to those in the ground state, i.e.,
$\Theta = \rho_{22} + \rho_{33}/\rho_{11}
$ which can be obtained by solving for the steadystate population of the three states using a standard master equation technique~\cite{PhysRevA.94.053859_seharris, PhysRevLett.66.1154_Imamo_3}. Expressions of \(\sigma_{abs}\), \(\sigma_{em}\) and $\Theta$
are well described in the literature~\cite{PhysRevLett.64.1107_Imamo_2, PhysRevA.94.053859_seharris, PhysRevLett.119.050602_zou, RevModPhys.77.633_Imamogl_1, PhysRevLett.66.1154_Imamo_3} and also provided in Supplementary Information 1. 
We consider a uniform pumping temperature $T_{13} = T_{23} = T_0$. The output of the system at the line center, \( B(\Delta\omega= 0) \), is intrinsically linked to the effective output radiation temperature \( T \) of the atomic system. This temperature can be calculated as:

\begin{equation}
    \label{eq:T_calc}
    T = \frac{\hbar \omega_{13}}{k \ln\left(\frac{1}{B(\Delta\omega= 0)} + 1\right)}
\end{equation}
where \( k \) is the Boltzmann constant. The normalized ratio \( T/T_0 \) compares the output radiation temperature \( T \) to the reservoir temperature \( T_0 \). Within the EIT regime, this temperature effectively assigns an energy value to the work mode of the system. However, \( T \) does not necessarily correspond to the maximum extractable work from the system, highlighting the need for a more detailed examination of \( T/T_0 \). To quantify the maximum extractable work the concept of ergotropy, \( \epsilon \) is employed~\cite{ccakmak2020ergotropy_coh, biswas2022extraction_erg}. Ergotropy provides a precise measure of the maximum work attainable through cyclic unitary transformations~\cite{van2020single_exp_erg, PhysRevA.110.032213_mjs_erg, ccakmak2020ergotropy_coh, biswas2022extraction_erg}. Experimental measurements of ergotropy have been performed in microscopic engines coupled to external loads~\cite{biswas2022extraction_erg, van2020single_exp_erg, elouard2023extending_exp_erg}. This metric enables the assessment of whether higher values of \( T/T_0 \) correspond to a greater potential for extractable work. It is interesting to note that work can only be extracted from an active state with respect to the system Hamiltonian. The concept is framed by considering two distinct density matrices: the active system density matrix \( \hat{\rho}(t) \) and a passive density matrix \( \hat{\rho}_P(t) \). Mathematically, the ergotropy is defined as the difference between the expectation value of the energy for the active density matrix and that for the passive density matrix:

\begin{equation}
    \label{ergotropy}
\epsilon = \langle \hat{H}_0 \hat{\rho}(t) \rangle - \langle \hat{H}_0 \hat{\rho}_P(t)\rangle,
\end{equation}

where \( \hat{H}_0 \) is the system Hamiltonian arranged in ascending eigenbasis, and \( \langle \hat{H}_0 \hat{\rho}_P(t)\rangle \) denotes the expectation value obtained from the passive density matrix. This passive state is specifically constructed to yield the minimum expectation value of the operator \( \hat{H}_0 \), ensuring that the difference in Eq. (\ref{ergotropy}) is maximal. The \textit{ergotropy}, $\epsilon$ for the three-level \(\Lambda\)-type system with the active state $diag(\rho_{22},\rho_{33},\rho_{11})$, is mathematically defined as:
\begin{align}
\epsilon = \hbar\omega_{23}\times(\rho_{33}-\rho_{22}).    
\end{align}

Through the evaluation of this quantity, we aim to identify an approach that enhances the performance of QHEs based on both \(T/T_0\) and \(\epsilon\). 

\section{Neural Network Mapping}
    \label{nn_mapp}

To develop an optimization protocol aimed at identifying high-performance engines based on the ratio \( T/T_0 \), we utilized an artificial neural network (ANN) to predict the upper states of an EIT-based three-level QHE. Accurately estimating the excited states is essential, as these states significantly influence the operational characteristics of the QHE, which are intrinsically linked to engine performance. The evaluation of engine performance involves several tunable parameters, including energy gaps between quantum states, ambient temperatures, laser characteristics, and Rabi frequencies. These parameters exhibit complex nonlinear interdependencies, complicating the determination of optimal configurations. Establishing a precise relationship between the ground and excited state properties—whether regarding emission dynamics or coupling strengths—often exceeds the capabilities of traditional analytical methods, thus necessitating advanced computational techniques.
In experimental setups for QHE design, the atomic system is first selected based on a suitable available protocol. For the chosen system, the primary tunable parameters are the Rabi frequency \( \Omega_C \) of the coupling laser and the surrounding thermal temperatures \( T_0 \). These parameters directly influence the system's behavior, particularly its capacity to leverage quantum coherences for enhanced performance. However, identifying the optimal combination of these parameters to achieve specific excited states poses significant experimental challenges due to the extensive parameter space involved. To predict states with elevated \( T/T_0 \) values, we begin with a ground state characterized by quantum numbers \( n_1, \ell_1, j_1 \), along with variables such as Rabi frequency, atomic and mass number, laser power, reservoir temperature, and effective output radiation temperature. Our objective is to ascertain whether predicting two additional states within this three-level system exhibiting higher \( T/T_0 \) values is possible. Another key motivation for this research is determining if higher \( T/T_0 \) ratios correlate with greater extractable work potential. To achieve this, we consider laser parameters including Rabi frequency \( \Omega_C \), transition frequencies \( \omega_{23} \) and \( \omega_{13} \), and set \( T_{13} = T_{23} = T_{0} \) as variable parameters. We then calculate \( T/T_0 \) using Eq. (\ref{eq:T_calc}) and aim to establish the following mapping:

\[
f: \{n_1, l_1, j_1, \Omega_C, P, T_0, T/T_0, Z, A\} \rightarrow \{n_2, l_2, j_2, n_3, l_3, j_3\}
\]

\begin{figure}[h]
\centering
    \includegraphics[width=0.4\textwidth]{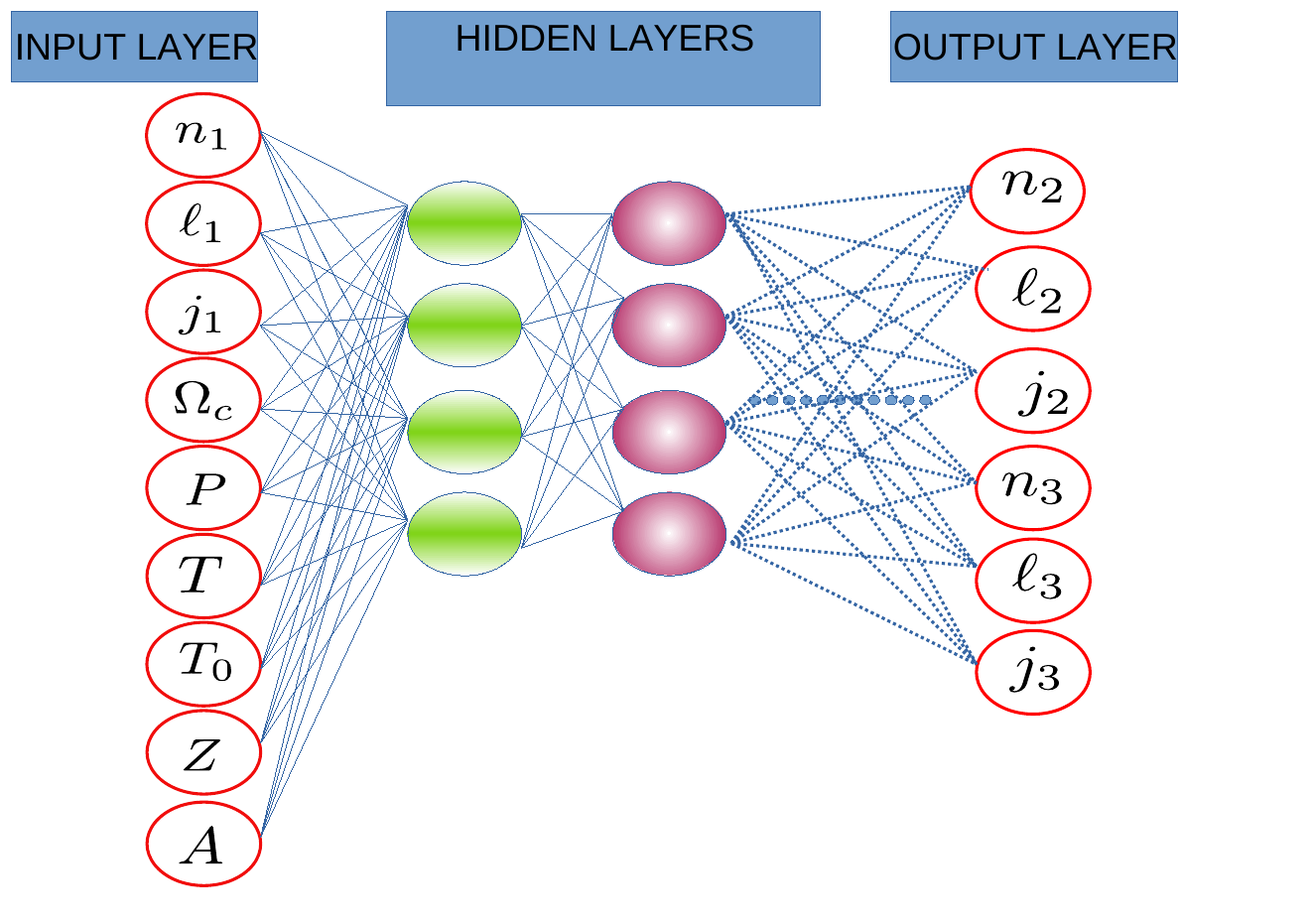}
    \caption{Schematic of the Artificial Neural Network Architecture. The input layer consists of 9 neurons with labels $n_1$, $\ell_1$, $j_1$, $\Omega_C$, $P$, $T$,$T_{13}=T_{23} = T_{0}$, $Z$ and $A$. The output layer consists of 6 neurons with labels $n_2$, $\ell_2$, $j_2$,  $n_3$, $\ell_3$ and $j_3$.}
    \label{fig:neural_network}
\end{figure}

To render Rabi frequency \( \Omega_C \), laser power \( P \), and system temperature \( T_0 \) dimensionless, we scale these parameters based on an experimentally observed engine~\cite{PhysRevLett.119.050602_zou} with reference values of \( \Omega_C = 10^8 \,\text{Hz} \), \( P = 130 \,\text{W} \), and \( T_0 = 5778 \,\text{K} \). Here, \( f \) denotes the learned many-to-many mapping between input and output parameters. The mapping is shown in Fig.~\ref{fig:neural_network}. The ANN model was trained using supervised learning technique to identify excited states across various QHE configurations.

\section{Data Generation and Modelling}
    \label{Modelling and Data Generation}
    
In order to generate the data we analyzed the alkali atoms Li, Na, K, Rb, Cs, and H and calculated the transition frequencies(\(\omega_{n,\ell,j,n',\ell',j'}\) in Hz), transition rates (\(R_{n,\ell,j,n',\ell',j'}\) in \(s^{-1}\)), and Rabi frequencies (\(\Omega_{C_{n,\ell,j,n',\ell',j'}}\) in Hz) for states characterized by the quantum number ranges as shown in Table (\ref{tab:data_table}) using the Alkali Rydberg Calculator (ARC) Python API \cite{vsibalic2017arc_ARC}. The laser polarization \( q \) was fixed at +1, and the laser waist was set to \( 50 \times 10^{-6} \, \text{m} \). The laser power \( P \) varied from 1 to 100 W, while the system temperature \( T_{0} \) ranged from 100 to 6000 K. The atomic number \( Z \) and mass number \( A \) were adjustable for defining the atom, and these parameters were used to calculate the effective temperature (\( T/T_0 \)) for each alkali metal atom using Eq. (\ref{eq:T_calc}). The quantum numbers \( n \), \( l \), and \( j \) were randomly generated under the constraint \( n_1 < n_2 < n_3 \). In total, we generated 4.6 million data points by varying \( n, l, j, \Omega_{C}, P, Z, A, \) and \( T_{0} \) within the ranges summarized in Table~\ref{tab:data_table}. It is important to note that we restricted the system to quantum numbers \( n \), \( l \), and \( j \) to quantify the levels of the engine, omitting the use of \( m_j \). The distribution of the generated data is provided in the Supplementary Information 3.


\begin{table*}
\caption{\label{tab:data_table}Range of Parameters for data generation.}
\begin{ruledtabular}
\begin{tabular}{cccccccc}
$n$ & $\ell$ & $j$ & $P$ & $\Omega_{C}$: via ARC & $T_{13}=T_{23} = T_{0}$ & $Z$ & $A$\\
\hline
$n_1$ $3$--$12$& $\ell_1$--$(1, 10)$ & $j_1$--$(0.5,10.5)$& ($1-130$) W ,7 values  & $300$ MHz--$24$ GHz& $100$--$6000$ K, $59$ values& 1 & \\
$n_2$ $4$--$13$& $\ell_2$--$(1, 10)$ & $j_2$--$(0.5,10.5)$& & & & 3& (6,7)\\
$n_3$ $6$--$14$& $\ell_3$--$(1, 11)$ & $j_3$--$(0.5,11.5)$ & & & & 11&\\
 &  & & & & & 19&(39,40,41)\\
 & &  & & & & 37&(85,87)\\
& & &  & & & 55&(133,137)\\

\end{tabular}
\end{ruledtabular}
\end{table*}

The quantum states and laser frequencies considered in this study are chosen to align with experimentally feasible configurations, ensuring the practical applicability of the theoretical framework. The energy levels and transitions of alkali atoms such as cesium, potassium, rubidium, sodium, lithium, and hydrogen are well-documented through high-resolution spectroscopy and laser cooling techniques~\cite{PhysRevA.98.052503, MPQ-274, PhysRevA.30.2881, JPhysB1976}. Hydrogen provides access to states including \(1s\), the metastable \(2s\), and Rydberg states (\(n \geq 10\))~\cite{PhysRevResearch.2.013244_H_atom_exp}. For cesium, experimentally accessible states include \(6S_{1/2}\), \(6P_{1/2,3/2}\), and Rydberg states (\(n \sim 18\))~\cite{PhysRevA.30.2881_all_alklai_exp_access}. Rubidium (\({85}\mathrm{Rb}\), \({87}\mathrm{Rb}\)) provides similar access to \(5S_{1/2}\), \(5P_{1/2,3/2}\), and Rydberg states (\(n \sim 50\))~\cite{PhysRevA.98.052503}. Potassium (\({39}\mathrm{K}, {40}\mathrm{K}, {41}\mathrm{K}\)), sodium, and lithium similarly exhibit ground, excited (\(nP_{1/2,3/2}\)), and Rydberg states (\(n \sim 20\))~\cite{PhysRevA.30.2881_all_alklai_exp_access}. These transitions correspond to wavelengths in the near-infrared to ultraviolet range~\cite{PhysRevA.98.052503, OE.17.015821}. Scalar polarizabilities of highly excited states, such as Rydberg states in cesium, have been precisely measured~\cite{JPhysB1976}, and modern techniques like cavity QED and magneto-optical traps enable control over state transitions and coherence properties~\cite{PhysRevResearch.2.013244}. Tunable lasers and high-precision spectroscopy ensure the laboratory implementation of the modeled laser-driven dynamics~\cite{PhysRevA.30.2881, OE.17.015821}.
We employed a systematic approach to determine the optimal amount of data required to train the ANN. Firstly, we use the 4.5 million initial dataset and shuffle it to ensure randomness. From this shuffled dataset, we select subsets containing 1000 to 500,000 data points and observe the performance of the neural network models based on two performance indicators: loss and mean absolute error (MAE) of the network.
The loss ($L$) for the ANN is defined as the mean squared error between the predicted output $\hat{y}$ and the true output $y$, given by \cite{michankow2024mean_loss_mae}:
\begin{equation*}
L = \frac{1}{N} \sum_{i=1}^{N} (\hat{y}_i - y_i)^2
\end{equation*}
where $N$ is the number of output data points. 

MAE for the ANN is defined as the average absolute difference between the predicted output $\hat{y}$ and the true output $y$, given by \cite{qi2020mean_MAE}:
\begin{equation*}
\text{MAE} = \frac{1}{N} \sum_{i=1}^{N} |\hat{y}_i - y_i|
\end{equation*}

Here, $\hat{y}_i$ represents the predicted output for the $i$-th input, and $y_i$ represents the true output for the $i$-th input. We split each subset into training and validation sets, allocating 80\% of the data for training and reserving 20\% for validation. Using TensorFlow's Keras API \cite{tensorflow2015-whitepaper}, we construct three ANN models, each featuring a single input layer with nine neurons, progressively increasing in complexity with two, three, and finally four hidden layers, culminating in an output layer with six neurons. We then perform manual hyperparameter tuning, iterating over various combinations of learning rates\cite{Goodfellow-et-al-2016_deep_learning}, number of neurons, and activation functions\cite{Seyrek2024_act_fns)} to optimize each model's performance. The model's performance is evaluated based on loss and MAE on the training and validation set for each subset of data. Finally, we visualize the ratio validation/training for both loss and the MAE wrt dataset size to assess the model's performance and identify potential overfitting or underfitting issues. All utilized resources and codes are available in the repository~\cite{githubGitHubManashSarmahDLEnabledPredictionofQHEsBasedonEIT}.

\section{ANN Results \label{ANNresults}}
\begin{figure}[h!]
  \centering
  \includegraphics[width= 8.5 cm, height = 12cm]{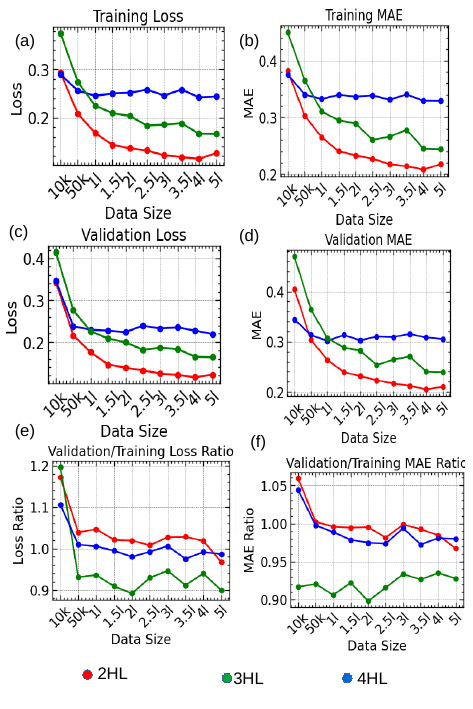}
  \caption{Performance metrics for ANNs with 2, 3, and 4 hidden layers across various dataset sizes. (a) Training Loss, (b) Training MAE, (c) Validation Loss, (d) Validation MAE, (e) Validation/Training Loss Ratio, (f) Validation/Training MAE Ratio. (In this context, "HL" stands for hidden layers, "k" means thousand, and "l" refers to lakh. ).}
  \label{fig:loss_mae_ratio_actual}
\end{figure}

In Figs.~\ref{fig:loss_mae_ratio_actual}a-f we plot the performance indicators as a function of the dataset size. Observing the performance indicators against varying dataset sizes up to 300 thousand samples, we note a saturation in both metrics, indicating no further increase in the performance of the models. We find that the ANN with two hidden layers (2HL) consistently outperforms its counterparts with three hidden layers (3HL) and four hidden layers (4HL) across various dataset sizes, as depicted in Figs.~\ref{fig:loss_mae_ratio_actual}(a-f). Specifically, the 2HL configuration achieves the most balanced performance in terms of both training and validation metrics. At larger dataset sizes, the 2HL network achieves a training loss of approximately 0.1250 and a training MAE of 0.2170. These metrics indicate that the 2HL network is highly efficient at learning the underlying patterns in the training data while avoiding overfitting. The validation loss and MAE further highlight the superior generalization capabilities of the 2HL network. At larger dataset sizes, the 2HL ANN maintains a validation loss close to its training loss, and the validation MAE stabilizes around 0.2170. This near parity between training and validation metrics demonstrates that the 2HL configuration avoids overfitting and generalizes well to unseen data. This corresponds to an approximate prediction accuracy of 78.30\%, showcasing the 2HL ANN's capability to model complex data while maintaining reliable generalization. In comparison, the 3HL and 4HL configurations show higher validation loss and MAE, suggesting poor generalization. The validation-to-training loss and MAE ratios provide additional insights into the model's performance. For the 2HL network, these ratios remain close to unity across dataset sizes, further confirming its balanced performance. Conversely, the 3HL and 4HL networks exhibit ratios consistently below unity, indicating underfitting and suboptimal performance. In Table (\ref{tab:hyperparameter_table}), we present the optimized hyperparameters and the validation loss and MAE for the three networks trained on the dataset containing 300,000 data points. Subsequently, we use the ANN model with 2 hidden layers with 128 neurons each, a learning rate of 0.01, the \texttt{tanh} activation function, achieving a loss of 0.1250 and a mean absolute error (MAE) of 0.2170, and use this ANN for further analysis.

\begin{table*}
\caption{\label{tab:hyperparameter_table}Optimized hyperparameters and validation Loss and MAE of the three ANN models at a dataset size of 300,000.}
\begin{ruledtabular}
\begin{tabular}{cccccc}
No. of hidden layers (HL) & Learning rate  & No of neurons & Activation function & $Loss_{vl}$ & $MAE_{vl}$ \\
& [0.01, 0.1] & [32, 64, 128]& [`relu', `tanh']& & 
\\
\hline
2& 0.01 & [128,128]& `tanh'& 0.1250& 0.2170\\
3& 0.01 & [32,128,32]& `relu' & 0.2327& 0.3092\\
4& 0.01 & [64,128,64,128]& `relu'& 0.2026& 0.2846\\

\end{tabular}
\end{ruledtabular}
\end{table*}

\section{Model Predictions and Application}
\label{predict}
\begin{figure}[h!]
    \centering
    \includegraphics[width= 8.5 cm, height = 12cm]{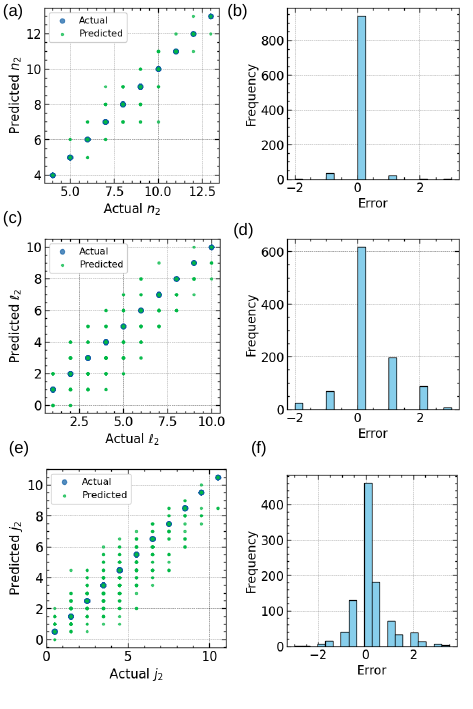}
    \caption{Actual vs. predicted values for $n_2$, $\ell_2$, and $j_2$ characterizing the second state. Panels (a), (c), and (e) show scatter plots of the actual versus predicted values for \(n_2\), \(l_2\), and \(j_2\) quantum numbers, respectively, with larger marker sizes representing higher data density. Panels (b), (d), and (f) present histograms of the prediction errors for \(n_2\), \(l_2\), and \(j_2\), respectively, indicating the model's accuracy by the concentration of errors around the zero value.}

    \label{fig:2state_pred}
\end{figure}

\begin{figure}[h!]
    \centering
    \includegraphics[width= 8.5 cm, height = 12cm]{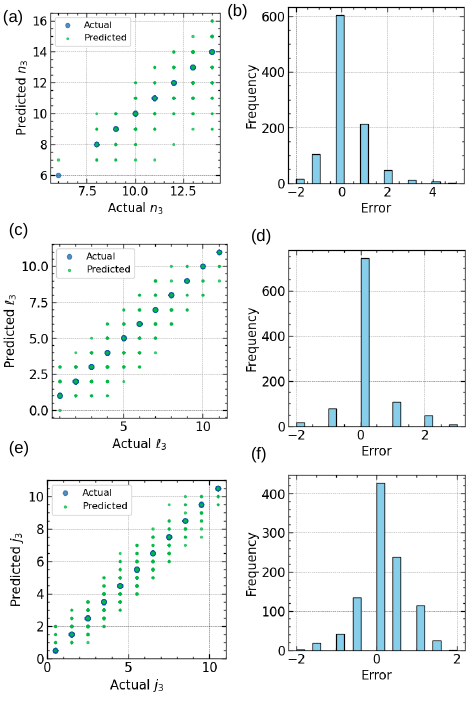}
    \caption{Actual vs. predicted values for $n_3$, $\ell_3$, and $j_3$ characterizing the third state. Panels (a), (c), and (e) show scatter plots of the actual versus predicted values for \(n_3\), \(l_3\), and \(j_3\) quantum numbers, respectively, with larger marker sizes representing higher data density. Panels (b), (d), and (f) present histograms of the prediction errors for \(n_3\), \(l_3\), and \(j_3\), respectively, indicating the model's accuracy by the concentration of errors around zero.}

    \label{fig:3state_pred}
\end{figure}

Using the trained model we start by predicting the excited states of alkali metal atoms for use in a QHE setup. From the initial dataset consisting of 45 million data points, we selected a subset of 1000 points for training and validation of the model. The predictions generated by our model and the associated error distributions are illustrated in Figs. \ref{fig:2state_pred} and \ref{fig:3state_pred}.  
Overall, the ANN model appears to perform reasonably well in predicting the quantum numbers $n$, $\ell$, and $j$, with lower errors observed for $n$ compared to $\ell$ and $j$. All error distributions are peaked around the zero value.

We filtered out certain QHEs with common states from the predicted data for all the considered atoms. These engines operate within three distinct output temperature regimes, characterized by \( T/T_0 \) values. These regimes are categorized according to the experimental framework outlined in Ref.~\cite{PhysRevLett.119.050602_zou}: the \textit{low range} corresponds to \( T/T_0 < 2.24 \), the \textit{mid-range} to \( 2.24 \leq T/T_0 \leq 3.0 \), and the \textit{high range} to \( T/T_0 > 3.0 \). The engines analyzed within each range involve specific transitions between atomic energy levels. In the \textit{low-output temperature regime}, the transitions are from ground states \( 10P_{2/3} \) to excited states \( 11P_{2/3} \) and \( 14D_{5/2} \). For the \textit{mid-output temperature regime}, the transitions occur between ground states \( 10F_{5/2} \) and excited states \( 11F_{5/2} \) and \( 14G_{7/2} \). In the \textit{high-output temperature regime}, the engines are characterized by transitions from ground states \( 8H_{9/2} \) to excited states \( 9F_{5/2} \) and \( 14G_{7/2} \). In the three regimes, we inspect whether \( T/T_0 \) corresponds to important thermodynamic quantities measuring the performance of the engines. We chose to evaluate the work (\( W \)) and ergotropy (\( \epsilon \)) for all the engines in the three regimes and attempt to draw a parallel between \( T/T_0 \), $W$, and $\epsilon$. The work done for the transition \( |1\rangle \rightarrow |3\rangle \) is given by:
\begin{equation}
    W = \Delta E - T \Delta S,
\end{equation}
where the energy difference is:$\Delta E = \hbar \omega_{13}$ and the entropy change associated with the transition is $\Delta S = -{\hbar \omega_{13}}/{T_0} - {\hbar \omega_{23}}/{T_0} - {\hbar \omega_{13}}/{T}$~\cite{PhysRevA.94.053859_seharris, PhysRevLett.119.050602_zou}. 



\begin{figure}[htbp]
    \centering
    \includegraphics[width= 8.5 cm, height = 12cm]{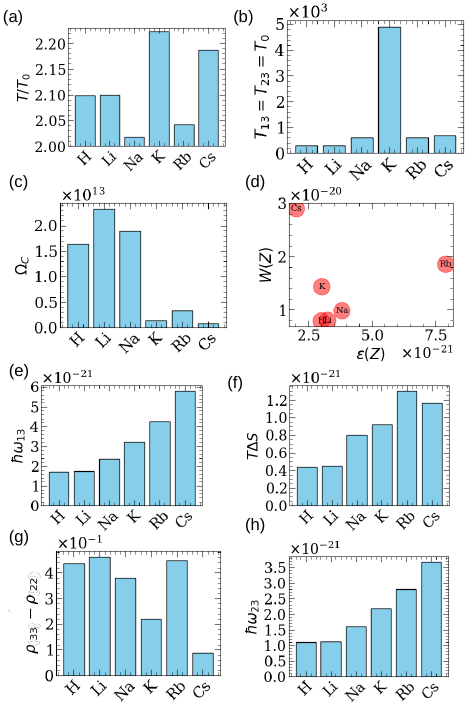}
        \caption{Comparison of various thermodynamic properties for alkali atoms (H, Li, Na, K, Rb, Cs) in a QHE setup for low range of \(T/T_0\). 
    (a) Ratio of the output radiation temperature to the bath temperature \( T/T_0 \). 
    (b) Temperature of the reservoirs \( T_{13} = T_{32} = T_0 \) in K.
    (c) Rabi frequency \(\Omega_C\) between states \(11P_{2/3}\) and \(14D_{5/2}\) in Hz.
    (d) Work, \( W(Z) \) in J versus ergotropy \( \epsilon(Z) \) in J for each atom as a function of the atomic number. 
    (e) Transition energy \(\hbar \omega_{13}\) between states \(10P_{3/2}\) and \(14D_{5/2}\) in J. 
    (f) Entropy change \(T\Delta S\) in J. 
    (g) Population difference between states, \(14D_{5/2}\) and \(11P_{2/3}\), \(\rho_{33} - \rho_{22}\). 
    (h) Transition energy \(\hbar \omega_{23}\) between states \(11P_{2/3}\) and \(14D_{5/2}\) in J. }
    \label{fig:low}
\end{figure}
\begin{figure}[htbp]
    \centering
    \includegraphics[width= 8.5 cm, height = 12cm]{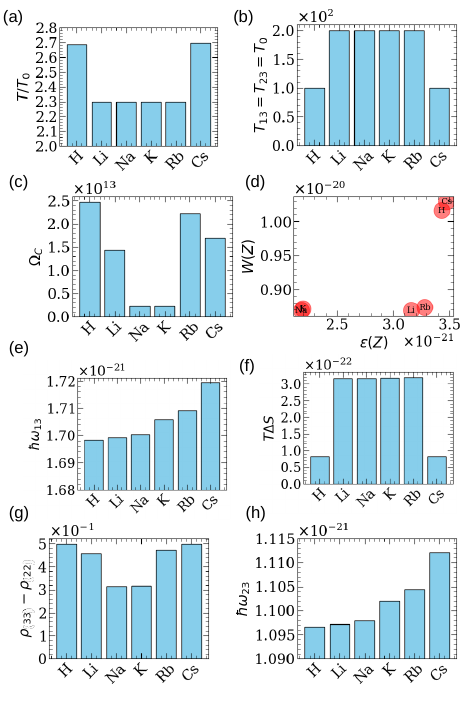}
        \caption{Comparison of various thermodynamic properties for alkali atoms (H, Li, Na, K, Rb, Cs) in a QHE setup for low range of \(T/T_0\). 
    (a) Ratio of the output radiation temperature to the bath temperature \( T/T_0 \). 
    (b) Temperature of the reservoirs \( T_{13} = T_{32} = T_0 \) in K.
    (c) Rabi frequency \(\Omega_C\) between states \(11F_{5/2}\) and \(10F_{5/2}\) in Hz.
    (d)  Work, \( W(Z) \) in J versus ergotropy \( \epsilon(Z) \) in J for each atom as a function of the atomic number. 
    (e) Transition energy \(\hbar \omega_{13}\) between states \(10F_{5/2}\) and \(14G_{7/2}\) in J. 
    (f) Entropy change \(T\Delta S\) in J. 
    (g) Population difference between states, states \(11F_{5/2}\) and \(10F_{5/2}\), \(\rho_{33} - \rho_{22}\). 
    (h) Transition energy \(\hbar \omega_{23}\) between states \(11F_{5/2}\) and \(10F_{5/2}\) in J. }
    \label{fig:mid}
\end{figure}

In Figs.~\ref{fig:low}(a-h), we analyze the engines based on \( T/T_0 \), \( W \), and \( \epsilon \) in the low-output temperature regime. In Figs.~\ref{fig:low}(a-c), we show the output radiation temperature \( T/T_0 \), reservoir temperature \( T_{13} = T_{23} = T_0 \), and \( \Omega_C \) values, respectively, that were used as input for the ANN. In Fig.~\ref{fig:low}(d), we present the variation of \( \epsilon \) and \( W \) for the different alkali metals. In Figs.~\ref{fig:low}(e,f), we show the variation of \( \hbar \omega_{13} \) and \( T\Delta S \), which help explain the observed work values. Finally, in Figs.~\ref{fig:low}(g,h), we display the variation of \( \rho_{33} - \rho_{22} \) and \( \hbar \omega_{23} \), which dictate the \( \epsilon \) values. In this regime, potassium (K) exhibits the highest \( T/T_0 \) value, followed by cesium (Cs) (Fig.~\ref{fig:low}(a)). However, potassium (K) does not produce the highest work (\( W \)) or ergotropy (\( \epsilon \)) (Fig.~\ref{fig:low}(d)). This is because \( \hbar \omega_{13} \), the energy gap, is relatively small for K (Fig.~\ref{fig:low}(e)), \( T \Delta S \), the entropy contribution, is relatively large (Fig.~\ref{fig:low}(f)), and the population difference \( \rho_{33} - \rho_{22} \) is also small (Fig.~\ref{fig:low}(g)). On the other hand, cesium (Cs) achieves the highest work (\( W \)) due to its large \( \hbar \omega_{13} \) (Fig.~\ref{fig:low}(e)) and relatively small \( T \Delta S \) (Fig.~\ref{fig:low}(f)). However, Cs has lower ergotropy (\( \epsilon \)) because the population difference \( \rho_{33} - \rho_{22} \) is small (Fig.~\ref{fig:low}(g)), even though it has a high \( \hbar \omega_{23} \) value (Fig.~\ref{fig:low}(h)). Finally, rubidium (Rb) exhibits the highest ergotropy (\( \epsilon \)) because it has both a large population difference \( \rho_{33} - \rho_{22} \) (Fig.~\ref{fig:low}(g)) and a high \( \hbar \omega_{23} \) value (Fig.~\ref{fig:low}(h)). 
In the mid-temperature regime, hydrogen (H) and cesium (Cs) exhibit elevated \( T/T_0 \) values (Fig.~\ref{fig:mid}(a)). Cesium achieves high work (\( W \)) due to its large energy gap (\( \hbar \omega_{13} \)), the highest among the elements considered (Fig.~\ref{fig:mid}(e)) and relatively low entropy contribution \( T\Delta S \) (Fig.~\ref{fig:mid}(f)). Similarly, hydrogen produces significant work, attributed to its comparable energy gap (\( \hbar \omega_{13} \)) (Fig.~\ref{fig:mid}(e)), its high \( T/T_0 \) value (Fig.~\ref{fig:mid}(a)), and relatively low entropy contribution \( T\Delta S \) (Fig.~\ref{fig:mid}(f)). The ergotropy (\( \epsilon \)) values for both H and Cs are also high, primarily due to their large population differences (\( \rho_{33} - \rho_{22} \)) (Fig.~\ref{fig:mid}(e)). Additionally, cesium benefits from a high \( \hbar \omega_{23} \) (Fig.~\ref{fig:mid}(h)), which further enhances its ergotropy. Hydrogen also has a similar \( \hbar \omega_{23} \) (Fig~\ref{fig:mid}(h)), its strong population difference (Fig~\ref{fig:mid}(g)) and low entropy contribution (Fig~\ref{fig:mid}(f)) allow it to maintain high ergotropy. 

\begin{figure}[htbp]
    \centering
    \includegraphics[width= 8.5 cm, height = 12cm]{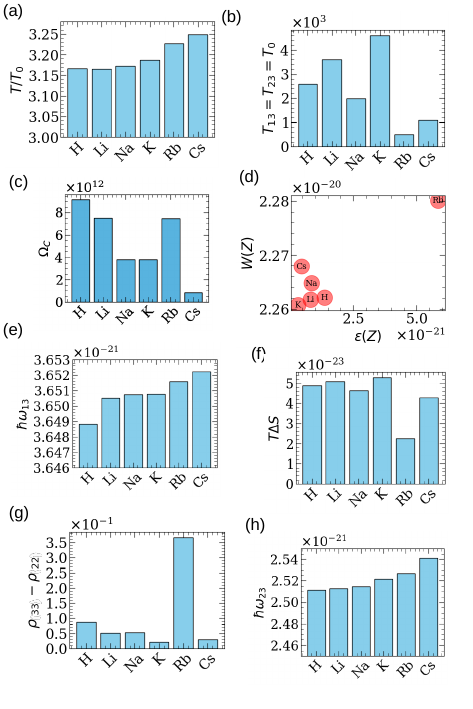}
        \caption{Comparison of various thermodynamic properties for alkali atoms (H, Li, Na, K, Rb, Cs) in a QHE setup for low range of \(T/T_0\). 
    (a) Ratio of the output radiation temperature to the bath temperature \( T/T_0 \). 
    (b) Temperature of the reservoirs \( T_{13} = T_{32} = T_0 \) in K.
    (c) Rabi frequency \(\Omega_C\) between states \(9F_{5/2}\) and \(14G_{7/2}\) in Hz.
    (d)  Work, \( W(Z) \) in J versus ergotropy \( \epsilon(Z) \) in J for each atom as a function of the atomic number.
    (e) Transition energy \(\hbar \omega_{13}\) between states \(8H_{9/2}\) and \(14G_{7/2}\) in J. 
    (f) Entropy change \(T\Delta S\) in J. 
    (g) Population difference, \(9F_{5/2}\) and \(14G_{7/2}\),between states \(\rho_{33} - \rho_{22}\). 
    (h) Transition energy \(\hbar \omega_{23}\) between states \(9F_{5/2}\) and \(14G_{7/2}\) in J.}
    \label{fig:high}
\end{figure}

In the high-output temperature regime, rubidium (Rb) and cesium (Cs) exhibit the highest \( T/T_0 \) values (Fig.~\ref{fig:high}(a)). Rubidium achieves superior work (\( W \)) and ergotropy (\( \epsilon \)) due to its large energy gap (\( \hbar \omega_{13} \)) (Fig.~\ref{fig:high}(c)), significant population difference (\( \rho_{33} - \rho_{22} \)) (Fig.~\ref{fig:high}(g)), and relatively low entropy contribution (\( T\Delta S \)) (Fig.~\ref{fig:high}(f)). In contrast, cesium (Cs), despite having a comparable \( T/T_0 \) value (Fig.~\ref{fig:high}(a)), exhibits lower work (\( W \)) and ergotropy (\( \epsilon \)) (Fig.~\ref{fig:high}(d)). This is primarily due to a slightly higher entropy contribution (\( T\Delta S \)) (Fig.~\ref{fig:high}(f)) and smaller population difference (\( \rho_{33} - \rho_{22} \)) (Fig.~\ref{fig:high}(e)). These findings demonstrate that \( T/T_0 \) alone is insufficient to determine engine performance in all regimes. While it can serve as a reliable metric in the mid-range, it fails in the low and high regimes. 

\begin{figure}[h!]
    \centering
    \includegraphics[width= 8 cm]{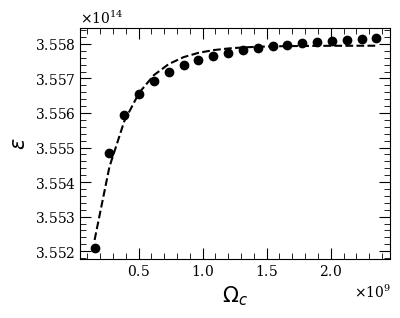}
    \caption{Ergotropy($\epsilon$) in Hz of $^{87}$Rb based QHE as a function of Rabi Frequency ($\Omega_C$) in Hz for $T/T_0 = 2$. The scatter points represent the actual data and the dotted line represents the fitted model, Eq~\ref{model_fitting}. }
    \label{fig:application}
\end{figure}

To further quantify if Rabi Frequency also plays an important more in determining ergotropy and work we explore the relation between Rabi frequency and ergotropy. To do this we analyze an Rb-based engine with ground-state configuration \( 6F_{7/2} \), where excited states \( 8P_{3/2} \) and \( {10}D_{5/2} \) correspond to \( T/T_0 = 2 \) and we plot the ergotropy till \(10^{9}Hz\). In Fig~\ref{fig:application}, we show the variation of \(\epsilon\) with \(\Omega_C\). The ergotropy as a function of \( \Omega_C \) exhibits a nonlinear behavior, characterized by a rapid increase followed by saturation.  At low \( \Omega_C \), the ergotropy is limited due to weak coupling. As \( \Omega_C \) increases, the system achieves greater population redistribution, maximizing the extractable work. However, the saturation of \( \epsilon \) at high \( \Omega_C \) indicates that there is an upper limit to the ergotropy, beyond which further driving yields no additional work and hence after a certain value of \(\Omega_C\) (\(\sim 10^9Hz\)) the ergotropy saturates and since each of \(\Omega_C\) for all three cases considered is greater than \(\sim 10^9Hz\) suggest that Rabi frequency has no role to play in the considered cases. This behavior is quantified using an exponential model:
\begin{equation}
    \epsilon(\Omega_C) = a (1 - e^{-b\Omega_C}) + c,
    \label{model_fitting}
\end{equation}

where parameters \( a , b , c \) are fitting constants. The best-fit parameters were determined as follows: 
\( a = 1.066\times10^{12}, b = 4.077\times10^{-9}, c = 3.547\times10^{14}\). The dotted curve illustrates our fitted model while scatter points reflect actual data observations. A similar trend is consistently observed across all alkali metal atoms. 

\begin{figure}[h]
    \centering
    \includegraphics[width= 8.5 cm, height = 12 cm]{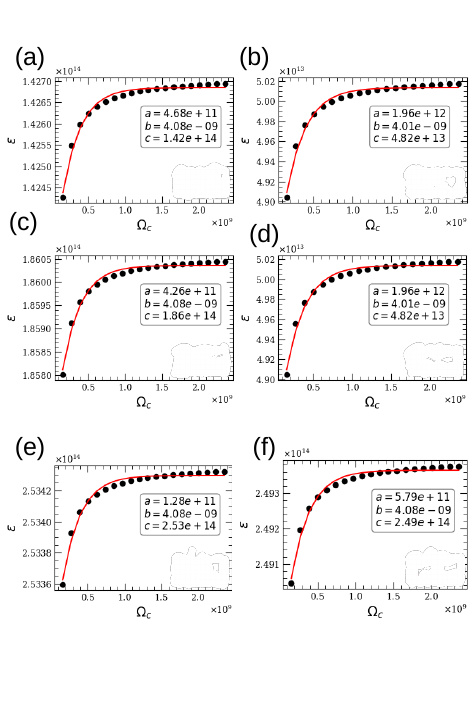}
        \caption{Ergotropy (\(\epsilon\)) in Hz as a function of Rabi Frequency (\(\Omega_C\)) in Hz of alkali atoms (H, Li, Na, K, Rb, Cs) with output radiation temperature (\(T/T_0\))$\approx$ 3. The scatter points represent the actual data and the red line represents the fitted model, Eq~\ref{model_fitting}.}
    \label{fig:app22}
\end{figure}

Fig.~\ref{fig:app22} illustrates the ergotropy ($\epsilon$) as a function of $\Omega_C$ for alkali atoms (H, Li, Na, K, Rb, Cs) at an output radiation temperature $T/T_0 \approx 3$. Each subplot corresponds to a specific atom, demonstrating the fitted curve (red line) based on Eq~\ref{model_fitting} alongside the numerical data (black dots). The fitting parameters $a$, $b$, and $c$ are presented within each plot. The results again reveal the saturation of ergotropy with increasing $\Omega_C$, highlighting the consistency of the fitted model with the numerical data. 

\section{Conclusion \label{conclusion}}

In this study, we developed an artificial neural network (ANN) model to predict the excited states of \( \Lambda \)-type alkali atom-based quantum heat engines (QHEs) which are of utmost importance to experimentalists. Such type of engines have been experimentally demonstrated within the working regime of electromagnetically induced transparency with cold atoms serving as the platform. The ANN algorithm successfully mapped the input domain spanning the ground state quantum numbers, Rabi frequency, laser power, reservoir temperature, atomic number, mass number, and the effective output radiation temperature to the desired output domain. We defined the output domain to be the excited states of the QHE. Following extensive hyperparameter tuning, the optimal neural network architecture comprised of two hidden layers with 128 neurons each, a learning rate of 0.01, and a `tanh' activation function. This configuration achieved a validation loss of 0.1250 and a validation mean absolute error (MAE) of 0.2170, corresponding to an approximate accuracy of 78.30\%. The analysis of QHEs across three distinct output radiation temperature regimes reveal that the engine performance needs to be assessed through the macroscopic work as well as ergotropy. The energy gap of states $|1\rangle$ and $|3\rangle$ along with the population difference between states $|2\rangle$ and $|3\rangle$ dictate the overall performance of the QHE through the three metrics. For the atomic engines with common states but different atomic numbers, the work was found not to increase with the increase in energy gap between states $|1\rangle$ and $|3\rangle$. This is due to entropic contributions in the work mode which is different for the different atoms and quantum numbers. The ergotropy, being a product of the energy difference and the population difference between states $|2\rangle$ and $|3\rangle$, also does not follow a strict trend with the increasing atomic number or quantum numbers. In the low-output temperature regime, although K was found to have the highest output radiation temperature, cesium demonstrated the maximum work output due to its relatively large energy gap coupled with a minimal entropy contribution. Meanwhile, rubidium possessed the highest ergotropy, primarily attributed to its significant population difference and a substantial energy gap. For the mid-output temperature regime, both hydrogen and cesium excelled in terms of work and ergotropy. Their superior performance was attributed to the respective high energy gaps and high population differences. In the high-output temperature regime, rubidium surpassed other alkali metals in both work and ergotropy due to its optimal energy gaps and low entropy contribution. Rb's dominance persisted despite cesium's comparable output radiation temperature. Our findings reveal that the output radiation temperature alone does not serve as an only predictor for engine performance. While the output radiation temperature shows a strong correlation with work and ergotropy in the mid-output temperature regime, additional factors, such as energy gaps, population differences, and entropy contributions, emerge as decisive in the low- and high-output temperature regimes. We also investigated the influence of Rabi frequency (\(\Omega_C\)) on ergotropy in an Rb-based engine. The ergotropy (\(\epsilon\)) was found to exhibit a nonlinear increase with \(\Omega_C\) due to enhanced population redistribution among states \(|2\rangle \) and \(|3\rangle \) , followed by saturation. This led us to conclude that \(\Omega_C\) has no significant role in improving ergotropy beyond a certain \(\Omega_C\) limit characteristic to the atom and quantum number under consideration. These results emphasize the multifaceted nature of QHE optimization and highlight the importance of a comprehensive parameter space analysis for enhancing the design and operation of quantum devices under varying thermodynamic conditions. Our study paved the way to integrate learning recipe with quantum mechanical data during the experimental prognosis of QHEs.
\begin{acknowledgments}
MS and HPG acknowledge the financial support from the Science and Engineering Research Board (Anusandhan National Research Foundation), India for the startup grant SRG/2021/001088.
\end{acknowledgments}

\section*{Supplementary Information 1}
The atomic configuration depicted in Fig.~(\ref{fig:QHE_levels}) has been previously analyzed \cite{PhysRevLett.64.1107_Imamo_2, RevModPhys.77.633_Imamogl_1, PhysRevLett.66.1154_Imamo_3} as a representative closed system applicable to lasers that operate without necessitating population inversion. The Hamiltonian of the system is given by~\cite{PhysRevLett.119.050602_zou, RevModPhys.77.633_Imamogl_1, PhysRevA.94.053859_seharris, NIAZ2024171816}:
 :

\[
\hat H = \hat H_0 + \hat H_\text{int},
\]

where:
\[
\hat H_0 = \hbar \omega_1 |1\rangle \langle 1| + \hbar \omega_2 |2\rangle \langle 2| + \hbar \omega_3 |3\rangle \langle 3|,
\]

and:
\[
\hat H_\text{int} = -\hbar \left( g e^{-i\omega t} |3\rangle \langle 1| + \Omega_C e^{-i\omega_c t} |3\rangle \langle 2| + \text{H.c.} \right).
\]
Here, \(g\) is the Rabi frequency associated with the radiating field that couples the states \(|1\rangle\) and \(|3\rangle\), quantifying the strength of the interaction between the radiating field and the system. \(\omega_c\) is the frequency of the control field (\(\Omega_C\)) driving the transition between \(|2\rangle\) and \(|3\rangle\). Because this system is closed, the driving blackbody excitation rates and the dephasing rates are determined by the lifetime decay rates \( \Gamma_{31} \) and \( \Gamma_{32} \), as well as the ambient temperatures. The transition rates \( R_{ij} = R_{ji} \) are linked to the thermal occupation numbers \( \bar{n}_{13} \) and \( \bar{n}_{23} \) as follows:

\begin{equation*}
    R_{23} = \Gamma_{32}\bar{n}_{23} = \Gamma_{32} \left\{\exp\left[\frac{\hbar\omega_{23}}{k_B T_{23}}\right] - 1\right\}^{-1},
\end{equation*}
\begin{equation*}
    R_{13} = \Gamma_{31}\bar{n}_{13} = \Gamma_{31} \left\{\exp\left[\frac{\hbar\omega_{13}}{k_B T_{13}}\right] - 1\right\}^{-1}.
\end{equation*}

The dephasing rates for each transition are given by, $ \gamma_{21} = R_{23} + R_{13}$, $\gamma_{31} = \Gamma_{31} + \Gamma_{32} + R_{23} + 2R_{13}$ and $ \gamma_{32} = \Gamma_{31} + \Gamma_{32} + R_{13} + 2R_{23}$.


Hence, after selecting an atomic system, the only adjustable parameters are the Rabi frequency ($\Omega_C$) of the coupling laser and the ambient pumping temperatures \( T_{13} \) and \( T_{23} \). Assuming no reflection or scattering into other modes, the spectral brightness \( B(\omega,z) \) for a single transverse mode can be computed as a function of distance \( z \). Assuming refractive indices close to unity, such that reflection can be neglected, the spectral brightness \( B(\delta, z) \) of the emission at the \( |3\rangle \) to \( |1\rangle \) transition is described by the following equation\cite{0-19-503437-6_b_w_deri_book}:
\begin{align*}
\frac{{dB(\delta, z)}}{{dz}} + N\left(\sigma_{\text{abs}}\rho_{11} - \sigma_{\text{em}}(\rho_{22} + \rho_{33})\right)B(\delta, z) &=\\ \sigma_{\text{em}}(\rho_{22} + \rho_{33}), \quad 
\end{align*}
with the boundary condition \( B(\delta, z = 0) = 0 \). Here, \( \delta = \omega_{13} - \omega \) represents the detuning from the on-resonance transition frequency \( \omega_{13} \), \( N \) is the atomic number density, and \( \rho_{ii} \) are the diagonal elements of the density matrix. The parameters \( \sigma_{\text{abs}} \) and \( \sigma_{\text{em}} \) denote the absorption and emission cross sections, respectively, for a weak probe beam on the \( |1\rangle \) to \( |3\rangle \) transition \cite{0-19-503437-6_b_w_deri_book}. At sufficiently large values of \( z \), where all relevant spectral components are absorbed, the spectral brightness reaches its maximum value, denoted as \( B_{\text{black}}(\omega) \). Consequently, the limiting brightness at each spectral component is:

\begin{align*}
B_{\text{black}}(\omega) = \frac{\Theta\sigma_{\text{em}}}{(\sigma_{\text{abs}}-\Theta\sigma_{\text{em}})} .
\end{align*}
Here, \( \Theta \) is defined as the ratio of atoms in the upper manifold to those in the ground state, i.e., \( \Theta = (\rho_{22} + \rho_{33})/\rho_{11} \). 
We assume that the field on the \( |1 \rangle \rightarrow |3 \rangle \) transition is sufficiently weak so that the populations are influenced primarily by the driving rates \( R_{ij} \) and the strong coupling field ( $\Omega_c$ ). By solving for \( \rho_{ii} \), one can obtain the value of \( \Theta \).

\begin{equation*}
\Theta = \frac{R_{13} \left[ 2\Omega_c^2 + \gamma_{32}(\Gamma_{32} + 2R_{23}) \right]}{(\Gamma_{31} + R_{13}) \left(\Omega_c^2 + \gamma_{32}R_{23} \right)}.
\end{equation*}

The absorptive and emissive cross sections are derived through the solution of density matrix equations, under the condition that the Rabi frequency associated with the coupling laser significantly exceeds all other Rabi frequencies within the system \cite{PhysRevLett.66.1154_Imamo_3}. Denoting $\omega$ as the frequency of the emitted spontaneous radiation and $\Delta = \omega_{13} - \omega$, these cross sections can be expressed as:
\begin{align*}
\frac{\sigma_{\text{abs}}}{\sigma_0} &= \frac{\gamma_{31} \left[ \gamma_{21} \Omega_c^2 + \gamma_{31} \left( \gamma_{21}^2 + 4\Delta \omega^2 \right) \right]}
{4 \Delta \omega^2 \left(-2\Omega_c^2 + \gamma_{21}^2 + \gamma_{31}^2 \right) + \left( \Omega_c^2 + \gamma_{21} \gamma_{31} \right)^2 + 16 \Delta \omega^4}.
\end{align*}

\begin{widetext}
\begin{align*}
\frac{\sigma_{\text{em}}}{\sigma_0} &= \frac{\gamma_{31} \Gamma_{32} \Omega_C^2\left( \Omega_c^2 + \gamma_{21} \gamma_{31} - 4\Delta \omega^2 \right) + \gamma_{31} \left[ \gamma_{21} \left( \Omega_c^2 + \gamma_{21} \gamma_{31} \right) + 4\gamma_{31} \Delta \omega^2 \right] \left( \Omega_c^2 + \gamma_{32} R_{23} \right)}
{\left[4 \Delta \omega^2 \left(-2\Omega_c^2 + \gamma_{21}^2 + \gamma_{31}^2 \right) + \left( \Omega_c^2 + \gamma_{21} \gamma_{31} \right)^2 + 16 \Delta \omega^4 \right] \left( \gamma_{32} \Gamma_{32} + 2\Omega_c^2 + 2\gamma_{32} R_{23} \right)}
\end{align*}
\end{widetext}

Both cross sections are standardized to $\sigma_0 = \frac{2\omega_{13}|\mu_{13}|^2}{\varepsilon_0 c \hbar \gamma_{13}}$, where $\mu_{13}$ represents the transition matrix element. We consider a uniform pumping temperature $T_{13} = T_{23} = T_0$ and calculate $B_{\text{black}}(\omega)$ accordingly. The brightness at the line center, i.e., $\Delta = 0$, is given by,
\[
B_{\text{black}}(0) = 
-\frac{\bar{n}_{13} \left[ \gamma_{21} \gamma_{32} \Gamma_{32} \bar{n}_{23} + (\gamma_{21} + \Gamma_{32}) \Omega_c^2 \right]}{\Gamma_{32} \bar{n}_{13} \Omega_c^2 - \gamma_{21} \left( \gamma_{32} \Gamma_{32} \bar{n}_{23} + \Omega_c^2 \right)}.
\]
A quantity called the normalized temperature $T/T_0$, which corresponds to the spectral brightness $B_{black}(0)$ can be calculated as
\begin{equation*}
T   = \frac{(\hbar\omega_{13}/k)}{\ln(\frac{1}{B_{black}(0)} + 1)}.    \label{eq:to_cal}
\end{equation*}
The coupling laser emits a photon along the $z$-axis at temperature $T_B$, sacrificing one photon in the process. Given that the entropy of a monochromatic laser remains constant regardless of photon exchange \cite{scovil1959three_qt_th, 10.1119/1.1842732_Mungan_th, PhysRevB.75.214304_qh_th_entropy}, the condition for overall entropy increase is  \cite{scovil1959three_qt_th, PhysRevB.75.214304_qh_th_entropy}:

\begin{equation*}
S = -\frac{\hbar\omega_{13}}{T_{13}} + \frac{\hbar\omega_{23}}{T_{23}} + \frac{\hbar\omega_{13}}{T_{B}} \ge 0. 
\end{equation*}

\begin{equation*}
T_{B}\le \frac{T_{13}T_{23}\omega_{13}}{T_{23}\omega_{13}-T_{13}\omega_{23}}.
\end{equation*}
In our case, $T_{13} = T_{23} = T_{0}$, hence $T_{B} = \left[\frac{\omega_{13}}{\omega_{13} - \omega_{23}}\right]T_{0} = \left[\frac{\omega_{13}}{\omega_{12}}\right]T_{0}$.
We have thus assumed that $B_{\text{black}}(\omega)$ belongs to the category of low-grade work and possesses the same entropy as a filtered thermal beam \cite{PhysRevB.75.214304_qh_th_entropy}. As in a standard unsaturated laser, the temperature $T_B$ will indefinitely increase and we can compare the efficiency of the EIT-based engine to the prototype engine developed by Scovil and Schulz-DuBois \cite{scovil1959three_qt_th} and Geusic et al. \cite{PhysRev.156.343_Geusic}.

\newpage
\section*{Supplementary Information 2}
\subsection{Artificial Neural Network (ANN) Architecture}

In this study, we employed a carefully designed Artificial Neural Network (ANN) model for predicting quantum state parameters. The architecture consists of two fully connected hidden layers, each containing 128 neurons. The choice of 128 neurons per layer strikes a balance between model complexity and computational efficiency.

\subsubsection{Activation Function}
For the hidden layers of our artificial neural network (ANN), we utilized the hyperbolic tangent (tanh) activation function. This function is mathematically defined as \cite{dubey2022activation_tanh_activation_fncs, sefiks5597_tanh}:

\[
\text{tanh}(z) = \frac{e^z - e^{-z}}{e^z + e^{-z}}
\]

where \(z\) is the pre-activation input, defined as:

\[
z = \sum_{i=1}^{n} w_i x_i + b,
\]

where \(x_i\) represents the inputs to the neuron (either the input features or outputs from the previous layer), \(w_i\) denotes the weights associated with each input \(x_i\), \(b\) is the bias term, which shifts the activation function, \(n\) is the total number of inputs to the neuron. The output of the activation function, \(\text{tanh}(z)\), introduces non-linearity by mapping the pre-activation value \(z\) to the range \([-1, 1]\). The derivative of the tanh function, which is needed for backpropagation to update the weights, is given by:

\[
\frac{d}{dz} \text{tanh}(z) = 1 - \text{tanh}^2(z).
\]

The smooth gradient helps avoid problems associated with vanishing gradients, particularly in comparison to the sigmoid activation function \cite{Szandała2021_act_fns}. The fact that \(\text{tanh}(z)\) is zero-centered also improves the dynamics of weight updates during training, as the gradients will tend to have more balanced updates. Furthermore, the bias term \( b_i \) is essential in ensuring that neurons are not overly constrained by the input data distribution, providing flexibility by enabling the neurons to activate (or not) based on learned patterns during the training process. Thus, the combination of weights, biases, and the non-linear activation function facilitates the network's ability to generalize and learn sophisticated features.





\subsubsection{Output Layer and Objective Function}

The output layer of the network consists of 6 neurons, each corresponding to one of the predicted quantum state parameters: \( n_2 \), \( l_2 \), \( j_2 \), \( n_3 \), \( l_3 \), and \( j_3 \). These parameters represent the quantum numbers of the excited states of the system being modeled. A **linear activation function** is used in the output layer neurons. The linear activation function is defined as:

\[
a_i = z_i,
\]

where \( z_i \) is the weighted sum of the inputs to the \( i \)-th neuron, and \( a_i \) is the corresponding output. 





\subsubsection{Optimizer and Learning Rate}

The model is trained using the **Adam optimizer**, which stands for Adaptive Moment Estimation\cite{10.1007/978-981-99-3432-4_7_adam_optimiser}. Adam is a popular optimization algorithm that combines the advantages of two other methods: the AdaGrad and RMSProp optimizers. The update rule for Adam includes both the first moment (mean) and second moment (variance) of the gradient, and is given by:

\[
\theta_t = \theta_{t-1} - \frac{\eta}{\sqrt{\hat{v}_t} + \epsilon} \hat{m}_t,
\]

where \( \theta_t \) represents the parameters (weights and biases) at iteration \( t \), \( \eta \) is the learning rate, \( \hat{m}_t \) is the estimate of the first moment (mean) of the gradient at time step \( t \), \( \hat{v}_t \) is the estimate of the second moment (uncentered variance) of the gradient, \( \epsilon \) is a small constant to prevent division by zero. The Adam optimizer is employed with a learning rate of \( 0.01 \). This learning rate determines the step size for each iteration of the weight updates. Adam's adaptive learning rate mechanism adjusts the learning rate for each parameter individually based on its gradient's magnitude, making the training process more efficient and robust. This adaptability is particularly beneficial when training on complex tasks, like predicting quantum state parameters, where gradients may vary significantly across different parameters. In summary, the model is trained to minimize the MSE between the predicted and actual quantum state parameters, with the Adam optimizer providing dynamic learning rate adjustment to improve convergence and model performance.
\newpage
\section*{Supplementary Information 3}

\begin{figure}[h]
    \centering
    \includegraphics[width= 8.5 cm, height = 12cm]{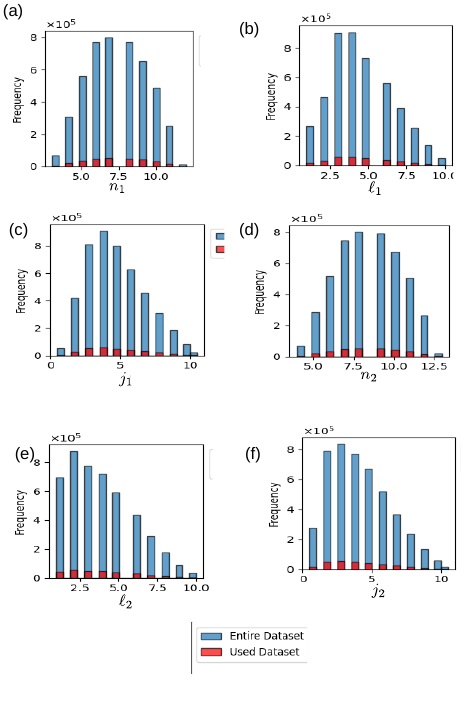}
    \caption{Histograms comparing the frequency distribution of quantum numbers \( n_1 \), \( l_1 \), \( j_1 \) and \( n_2 \), \( l_2 \), \( j_2 \) between the entire dataset (blue) and the used dataset (red). Panels (a) to (f) correspond to \( n_3 \), \( l_3 \), \( j_3 \), and \( n_2 \), \( l_2 \), \( j_2 \), respectively. The distributions demonstrate that the used dataset, although significantly smaller, follows similar trends as the entire dataset across all parameters.}
    \label{fig:quantum_distribution}
\end{figure}

The Fig.~\ref{fig:quantum_distribution} presents histograms that compare the distribution of various quantum numbers \( n_1 \), \( l_1 \), \( j_1 \), \( n_2 \), \( l_2 \), and \( j_2 \) between the entire dataset and the subset of data used in our analysis.

\begin{figure}[h!]
    \centering
    \includegraphics[width= 8.5 cm, height = 12cm]{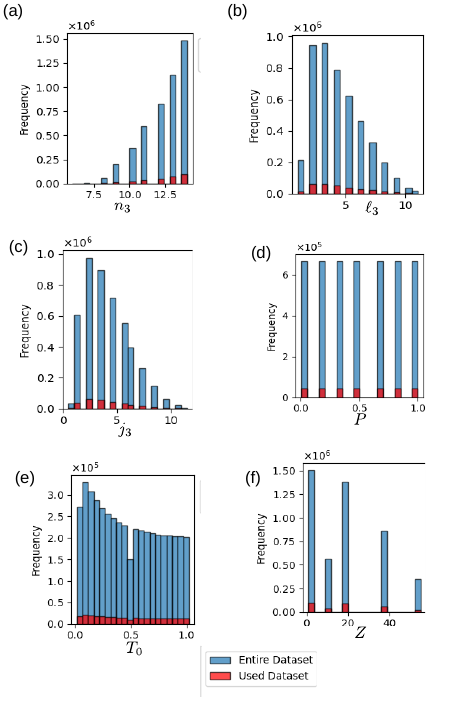}
    \caption{Histograms comparing the frequency distribution of quantum numbers \( n_3 \), \( l_3 \), \( j_3 \), laser power \( P \), reservoir temperature \( Te \), and atomic number \( Z \) between the entire dataset (blue) and the used dataset (red). Panels (a) to (f) correspond to \( n_3 \), \( l_3 \), \( j_3 \), \( LP \), \( Te \), and \( Z \), respectively. The distributions demonstrate that the used dataset, although significantly smaller, follows similar trends as the entire dataset across all parameters.}
    \label{fig:quantum_distribution_2}
\end{figure}

In panel (a), the frequency distribution of the quantum number \( n_1 \) shows that the most frequent values lie between 6.5 and 8.5 for the entire dataset, while the used dataset follows a similar trend, though at a much smaller scale. Panel (b) displays the distribution of the quantum number \( l_1 \), where the entire dataset peaks near \( l_1 = 5 \), and the used dataset exhibits a proportionally smaller but consistent frequency across the same range. Panel (c) represents the distribution for \( j_1 \), which shows a peak around \( j_1 = 5.5 \), with the used dataset being a small fraction of the total data. Panel (d) shows the frequency distribution of the second quantum number \( n_2 \), where the entire dataset peaks near \( n_2 = 7.5 \), and the used dataset displays a similar trend. In panel (e), the distribution for \( l_2 \) follows a decreasing trend as \( l_2 \) increases, with both datasets showing a peak near \( l_2 = 5 \). Finally, panel (f) presents the distribution for \( j_2 \), which also peaks around \( j_2 = 5.5 \), similar to the pattern observed for \( j_1 \). Across all panels, the blue bars (entire dataset) represent a much larger number of data points compared to the red bars (used dataset), emphasizing that only a small fraction of the total data was used for the analysis. Despite this, the used dataset retains the general trends observed in the entire dataset. The Fig.~\ref{fig:quantum_distribution_2} presents histograms comparing the frequency distributions of the quantum numbers \( n_3 \), \( l_3 \), \( j_3 \), as well as other relevant parameters such as laser power \( P \), reservoir temperature \( T_0 \), and atomic number \( Z \) between the entire dataset and the subset of data used in our analysis.
In panel (a), the frequency distribution of \( n_3 \) shows that the most frequent values are concentrated between 7.5 and 12.5 for the entire dataset, with the used dataset following the same trend but with fewer data points. Panel (b) depicts the distribution of the quantum number \( l_3 \), where the entire dataset shows a decreasing frequency as \( l_3 \) increases, with a peak at lower values. The used dataset also follows this pattern but at a reduced scale. In panel (c), the distribution of \( j_3 \) peaks around 5, with both datasets displaying a similar trend. Panel (d) shows the distribution for laser power \( P \). The distributions for reservoir temperature \( T_0 \) in panel (e) show a roughly exponential decrease, with the entire dataset peaking near lower values of \( T_0 \). The used dataset is a smaller subset but mirrors the same trend. Finally, panel (f) presents the atomic number \( Z \), where the frequency distribution shows distinct peaks for the entire dataset, particularly at \( Z = 1 \), \( Z = 19 \), and \( Z = 37 \), with the used dataset reflecting similar peaks.

\begin{figure}[h!]
        \centering
        \includegraphics[width= 8.5 cm, height = 10cm]{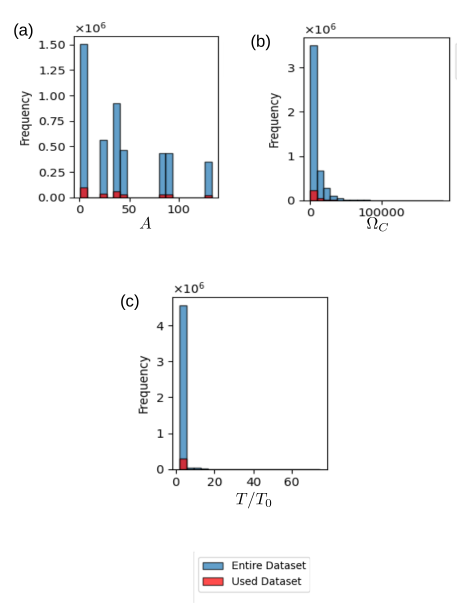}
\caption{Histograms comparing the frequency distribution of temperature, mass number, and Rabi frequency between the entire dataset (blue) and the used dataset (red). Panels (a) to (c) correspond to mass number,Rabi frequency and output radiation temperature, respectively. The entire dataset shows a significantly higher number of points, while the used dataset retains the general trends, focusing on a smaller subset of the total data. The distributions suggest similar patterns between the two datasets, with peaks concentrated near specific values for each parameter.}

        \label{fig:quantum_distribution_3}
\end{figure}    

\begin{figure}[h!]
    \centering
    \includegraphics[width= 8.5 cm, height = 12cm]{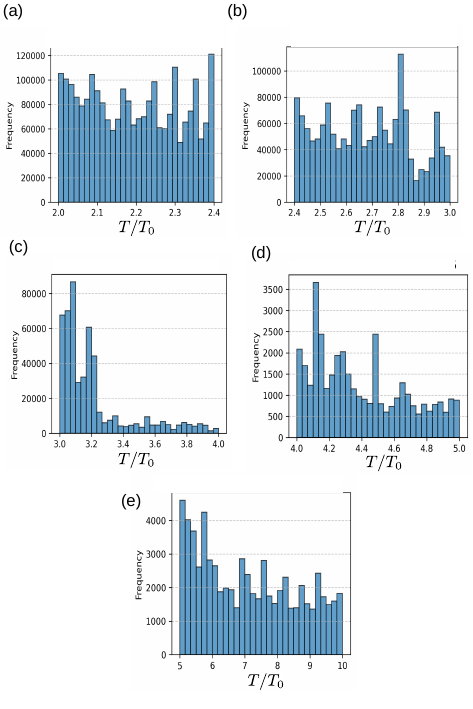}
    \caption{Histograms showing the distribution of \( T/T_0 \) values for lower ranges:
        (a) \( T/T_0 \) in the range [0, 2.4], showing high-frequency bins with consistent variability,
        (b) \( T/T_0 \) in the range [2.4, 3], highlighting pronounced peaks around \( T/T_0 = 2.8 \),
        (c) \( T/T_0 \) in the range [3, 4], showing a steep drop in frequency after \( T/T_0 = 3.2 \),
        (d) \( T/T_0 \) in the range [4, 5], revealing multiple moderate-frequency peaks,
        (e) \( T/T_0 \) in the range [5, 10], showing a decline in frequency as \( T/T_0 \) increases.}
    \label{fig:T_dis1}
\end{figure}

\begin{figure}[h!]
    \centering
    \includegraphics[width= 8.5 cm, height = 12cm]{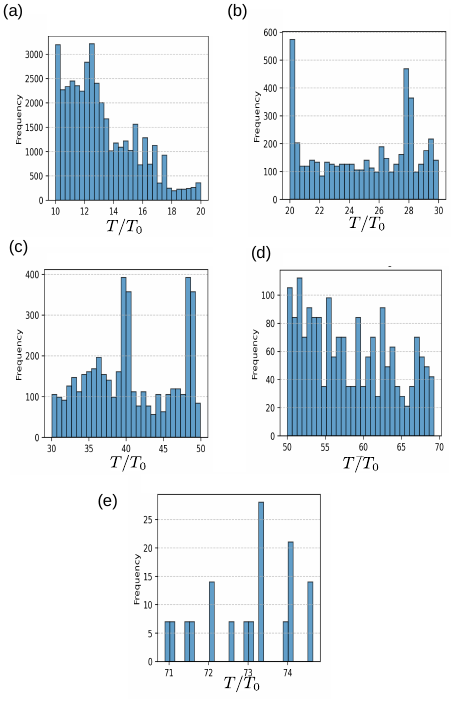}
    \caption{Histograms illustrating the distribution of \( T/T_0 \) values across various ranges: 
        (a) \( T/T_0 \) in the range [10, 20], showing a peak near \( T/T_0 = 12 \),
        (b) \( T/T_0 \) in the range [20, 30], indicating multiple peaks across the range,
        (c) \( T/T_0 \) in the range [30, 50], revealing clustered peaks at \( T/T_0 \approx 40 \) and \( T/T_0 \approx 45 \),
        (d) \( T/T_0 \) in the range [50, 70], highlighting spread-out peaks,
        (e) \( T/T_0 \) in the range [70, 74.723], showing sparse peaks with reduced frequency.}
    \label{fig:T_dis2}
\end{figure}

The blue bars represent the entire dataset, while the red bars show the data used for the analysis. Although the used dataset is a smaller subset, it follows the general trends observed in the entire dataset, ensuring that the subset is representative of the overall data. The Fig.~\ref{fig:quantum_distribution_3} presents histograms comparing the frequency distributions of the output radiation temperature \( T/T_0 \), mass number \( A \), and Rabi Frequency \( \Omega_C \), between the entire dataset and the subset of data used in our analysis. Fig.~\ref{fig:T_dis1} and Fig.~\ref{fig:T_dis2} we show the $T/T_0$ in different sections.




\bibliography{bibliograpgy}

\begin{thebibliography}{50}%
\makeatletter
\providecommand \@ifxundefined [1]{%
 \@ifx{#1\undefined}
}%
\providecommand \@ifnum [1]{%
 \ifnum #1\expandafter \@firstoftwo
 \else \expandafter \@secondoftwo
 \fi
}%
\providecommand \@ifx [1]{%
 \ifx #1\expandafter \@firstoftwo
 \else \expandafter \@secondoftwo
 \fi
}%
\providecommand \natexlab [1]{#1}%
\providecommand \enquote  [1]{``#1''}%
\providecommand \bibnamefont  [1]{#1}%
\providecommand \bibfnamefont [1]{#1}%
\providecommand \citenamefont [1]{#1}%
\providecommand \href@noop [0]{\@secondoftwo}%
\providecommand \href [0]{\begingroup \@sanitize@url \@href}%
\providecommand \@href[1]{\@@startlink{#1}\@@href}%
\providecommand \@@href[1]{\endgroup#1\@@endlink}%
\providecommand \@sanitize@url [0]{\catcode `\\12\catcode `\$12\catcode `\&12\catcode `\#12\catcode `\^12\catcode `\_12\catcode `\%12\relax}%
\providecommand \@@startlink[1]{}%
\providecommand \@@endlink[0]{}%
\providecommand \url  [0]{\begingroup\@sanitize@url \@url }%
\providecommand \@url [1]{\endgroup\@href {#1}{\urlprefix }}%
\providecommand \urlprefix  [0]{URL }%
\providecommand \Eprint [0]{\href }%
\providecommand \doibase [0]{http://dx.doi.org/}%
\providecommand \selectlanguage [0]{\@gobble}%
\providecommand \bibinfo  [0]{\@secondoftwo}%
\providecommand \bibfield  [0]{\@secondoftwo}%
\providecommand \translation [1]{[#1]}%
\providecommand \BibitemOpen [0]{}%
\providecommand \bibitemStop [0]{}%
\providecommand \bibitemNoStop [0]{.\EOS\space}%
\providecommand \EOS [0]{\spacefactor3000\relax}%
\providecommand \BibitemShut  [1]{\csname bibitem#1\endcsname}%
\let\auto@bib@innerbib\@empty
\bibitem [{sco(1959)}]{scovil1959three_qt_th}%
  \BibitemOpen
  \bibfield  {title} {\enquote {\bibinfo {title} {Three-level masers as heat engines},}\ }\href@noop {} {\bibfield  {journal} {\bibinfo  {journal} {Physical Review Letters}\ }\textbf {\bibinfo {volume} {2}},\ \bibinfo {pages} {262} (\bibinfo {year} {1959})}\BibitemShut {NoStop}%
\bibitem [{\citenamefont {Scovil}\ and\ \citenamefont {Schulz-DuBois}(1959)}]{PhysRevLett.2.262_scovil_1st}%
  \BibitemOpen
  \bibfield  {author} {\bibinfo {author} {\bibfnamefont {H.~E.~D.}\ \bibnamefont {Scovil}}\ and\ \bibinfo {author} {\bibfnamefont {E.~O.}\ \bibnamefont {Schulz-DuBois}},\ }\bibfield  {title} {\enquote {\bibinfo {title} {Three-level masers as heat engines},}\ }\href {\doibase 10.1103/PhysRevLett.2.262} {\bibfield  {journal} {\bibinfo  {journal} {Phys. Rev. Lett.}\ }\textbf {\bibinfo {volume} {2}},\ \bibinfo {pages} {262--263} (\bibinfo {year} {1959})}\BibitemShut {NoStop}%
\bibitem [{\citenamefont {Goswami}\ and\ \citenamefont {Harbola}(2013)}]{PhysRevA.88.013842_hpg01}%
  \BibitemOpen
  \bibfield  {author} {\bibinfo {author} {\bibfnamefont {H.~P.}\ \bibnamefont {Goswami}}\ and\ \bibinfo {author} {\bibfnamefont {U.}~\bibnamefont {Harbola}},\ }\bibfield  {title} {\enquote {\bibinfo {title} {Thermodynamics of quantum heat engines},}\ }\href {\doibase 10.1103/PhysRevA.88.013842} {\bibfield  {journal} {\bibinfo  {journal} {Phys. Rev. A}\ }\textbf {\bibinfo {volume} {88}},\ \bibinfo {pages} {013842} (\bibinfo {year} {2013})}\BibitemShut {NoStop}%
\bibitem [{\citenamefont {Xu}\ \emph {et~al.}(2024)\citenamefont {Xu}, \citenamefont {Jin}, \citenamefont {de~Almeida},\ and\ \citenamefont {Neto}}]{PhysRevB.110.134318_2_lev_2024}%
  \BibitemOpen
  \bibfield  {author} {\bibinfo {author} {\bibfnamefont {H.-G.}\ \bibnamefont {Xu}}, \bibinfo {author} {\bibfnamefont {J.}~\bibnamefont {Jin}}, \bibinfo {author} {\bibfnamefont {N.~G.}\ \bibnamefont {de~Almeida}}, \ and\ \bibinfo {author} {\bibfnamefont {G.~D. d.~M.}\ \bibnamefont {Neto}},\ }\bibfield  {title} {\enquote {\bibinfo {title} {Exploring the role of criticality in the quantum otto cycle fueled by the anisotropic quantum rabi-stark model},}\ }\href {\doibase 10.1103/PhysRevB.110.134318} {\bibfield  {journal} {\bibinfo  {journal} {Phys. Rev. B}\ }\textbf {\bibinfo {volume} {110}},\ \bibinfo {pages} {134318} (\bibinfo {year} {2024})}\BibitemShut {NoStop}%
\bibitem [{\citenamefont {xiang Deng}\ \emph {et~al.}(2024)\citenamefont {xiang Deng}, \citenamefont {He}, \citenamefont {Liu}, \citenamefont {Shao},\ and\ \citenamefont {Cui}}]{deng2024capturingdynamicsthermodynamicsthreelevel_3_lev_2024}%
  \BibitemOpen
  \bibfield  {author} {\bibinfo {author} {\bibfnamefont {G.}~\bibnamefont {xiang Deng}}, \bibinfo {author} {\bibfnamefont {Z.}~\bibnamefont {He}}, \bibinfo {author} {\bibfnamefont {Y.}~\bibnamefont {Liu}}, \bibinfo {author} {\bibfnamefont {W.}~\bibnamefont {Shao}}, \ and\ \bibinfo {author} {\bibfnamefont {Z.}~\bibnamefont {Cui}},\ }\href {https://arxiv.org/abs/2405.17763} {\enquote {\bibinfo {title} {Capturing dynamics and thermodynamics of a three-level quantum heat engine via programmable quantum circuits},}\ } (\bibinfo {year} {2024}),\ \Eprint {http://arxiv.org/abs/2405.17763} {arXiv:2405.17763 [quant-ph]} \BibitemShut {NoStop}%
\bibitem [{\citenamefont {Sandilya}\ \emph {et~al.}(2024)\citenamefont {Sandilya}, \citenamefont {Akhtar}, \citenamefont {Sarmah},\ and\ \citenamefont {Goswami}}]{https://doi.org/10.1002/andp.202400143_mriga}%
  \BibitemOpen
  \bibfield  {author} {\bibinfo {author} {\bibfnamefont {M.}~\bibnamefont {Sandilya}}, \bibinfo {author} {\bibfnamefont {J.}~\bibnamefont {Akhtar}}, \bibinfo {author} {\bibfnamefont {M.~J.}\ \bibnamefont {Sarmah}}, \ and\ \bibinfo {author} {\bibfnamefont {H.~P.}\ \bibnamefont {Goswami}},\ }\bibfield  {title} {\enquote {\bibinfo {title} {Cotunneling effects in the geometric statistics of a nonequilibrium spin-resolved junction},}\ }\href {\doibase https://doi.org/10.1002/andp.202400143} {\bibfield  {journal} {\bibinfo  {journal} {Annalen der Physik}\ }\textbf {\bibinfo {volume} {536}},\ \bibinfo {pages} {2400143} (\bibinfo {year} {2024})}\BibitemShut {NoStop}%
\bibitem [{\citenamefont {Leitch}\ \emph {et~al.}(2022)\citenamefont {Leitch}, \citenamefont {Piccione}, \citenamefont {Bellomo},\ and\ \citenamefont {De~Chiara}}]{10.1116/5.0072067_HO_2022}%
  \BibitemOpen
  \bibfield  {author} {\bibinfo {author} {\bibfnamefont {H.}~\bibnamefont {Leitch}}, \bibinfo {author} {\bibfnamefont {N.}~\bibnamefont {Piccione}}, \bibinfo {author} {\bibfnamefont {B.}~\bibnamefont {Bellomo}}, \ and\ \bibinfo {author} {\bibfnamefont {G.}~\bibnamefont {De~Chiara}},\ }\bibfield  {title} {\enquote {\bibinfo {title} {Driven quantum harmonic oscillators: A working medium for thermal machines},}\ }\href {\doibase 10.1116/5.0072067} {\bibfield  {journal} {\bibinfo  {journal} {AVS Quantum Science}\ }\textbf {\bibinfo {volume} {4}},\ \bibinfo {pages} {012001} (\bibinfo {year} {2022})},\ \Eprint {http://arxiv.org/abs/https://pubs.aip.org/avs/aqs/article-pdf/doi/10.1116/5.0072067/19803495/012001\_1\_online.pdf} {https://pubs.aip.org/avs/aqs/article-pdf/doi/10.1116/5.0072067/19803495/012001\_1\_online.pdf} \BibitemShut {NoStop}%
\bibitem [{\citenamefont {Sarmah}\ and\ \citenamefont {Goswami}(2024{\natexlab{a}})}]{PhysRevA.110.052214_mjs_efficiency}%
  \BibitemOpen
  \bibfield  {author} {\bibinfo {author} {\bibfnamefont {M.~J.}\ \bibnamefont {Sarmah}}\ and\ \bibinfo {author} {\bibfnamefont {H.~P.}\ \bibnamefont {Goswami}},\ }\bibfield  {title} {\enquote {\bibinfo {title} {Efficiency fluctuations of a heat engine with noise-induced quantum coherences},}\ }\href {\doibase 10.1103/PhysRevA.110.052214} {\bibfield  {journal} {\bibinfo  {journal} {Phys. Rev. A}\ }\textbf {\bibinfo {volume} {110}},\ \bibinfo {pages} {052214} (\bibinfo {year} {2024}{\natexlab{a}})}\BibitemShut {NoStop}%
\bibitem [{\citenamefont {Hegde}\ \emph {et~al.}(2023)\citenamefont {Hegde}, \citenamefont {Ott}, \citenamefont {Xia}, \citenamefont {Kasper}, \citenamefont {Berges},\ and\ \citenamefont {Jendrzejewski}}]{PhysRevA.107.L031302_fluctuations_exp}%
  \BibitemOpen
  \bibfield  {author} {\bibinfo {author} {\bibfnamefont {A.}~\bibnamefont {Hegde}}, \bibinfo {author} {\bibfnamefont {R.}~\bibnamefont {Ott}}, \bibinfo {author} {\bibfnamefont {A.}~\bibnamefont {Xia}}, \bibinfo {author} {\bibfnamefont {V.}~\bibnamefont {Kasper}}, \bibinfo {author} {\bibfnamefont {J.}~\bibnamefont {Berges}}, \ and\ \bibinfo {author} {\bibfnamefont {F.}~\bibnamefont {Jendrzejewski}},\ }\bibfield  {title} {\enquote {\bibinfo {title} {Nonequilibrium dynamics of fluctuations in an ultracold atomic mixture},}\ }\href {\doibase 10.1103/PhysRevA.107.L031302} {\bibfield  {journal} {\bibinfo  {journal} {Phys. Rev. A}\ }\textbf {\bibinfo {volume} {107}},\ \bibinfo {pages} {L031302} (\bibinfo {year} {2023})}\BibitemShut {NoStop}%
\bibitem [{\citenamefont {Sarmah}\ and\ \citenamefont {Goswami}(2023)}]{SARMAH2023129135}%
  \BibitemOpen
  \bibfield  {author} {\bibinfo {author} {\bibfnamefont {M.~J.}\ \bibnamefont {Sarmah}}\ and\ \bibinfo {author} {\bibfnamefont {H.~P.}\ \bibnamefont {Goswami}},\ }\bibfield  {title} {\enquote {\bibinfo {title} {Learning coherences from nonequilibrium fluctuations in a quantum heat engine},}\ }\href {\doibase https://doi.org/10.1016/j.physa.2023.129135} {\bibfield  {journal} {\bibinfo  {journal} {Physica A: Statistical Mechanics and its Applications}\ }\textbf {\bibinfo {volume} {627}},\ \bibinfo {pages} {129135} (\bibinfo {year} {2023})}\BibitemShut {NoStop}%
\bibitem [{\citenamefont {Sarmah}\ and\ \citenamefont {Goswami}(2024{\natexlab{b}})}]{PhysRevA.110.032213_mjs_ergotropy}%
  \BibitemOpen
  \bibfield  {author} {\bibinfo {author} {\bibfnamefont {M.~J.}\ \bibnamefont {Sarmah}}\ and\ \bibinfo {author} {\bibfnamefont {H.~P.}\ \bibnamefont {Goswami}},\ }\bibfield  {title} {\enquote {\bibinfo {title} {Noise-induced coherent ergotropy of a quantum heat engine},}\ }\href {\doibase 10.1103/PhysRevA.110.032213} {\bibfield  {journal} {\bibinfo  {journal} {Phys. Rev. A}\ }\textbf {\bibinfo {volume} {110}},\ \bibinfo {pages} {032213} (\bibinfo {year} {2024}{\natexlab{b}})}\BibitemShut {NoStop}%
\bibitem [{\citenamefont {Van~Horne}\ \emph {et~al.}(2020)\citenamefont {Van~Horne}, \citenamefont {Yum}, \citenamefont {Dutta}, \citenamefont {H{\"a}nggi}, \citenamefont {Gong}, \citenamefont {Poletti},\ and\ \citenamefont {Mukherjee}}]{van2020single_exp_erg}%
  \BibitemOpen
  \bibfield  {author} {\bibinfo {author} {\bibfnamefont {N.}~\bibnamefont {Van~Horne}}, \bibinfo {author} {\bibfnamefont {D.}~\bibnamefont {Yum}}, \bibinfo {author} {\bibfnamefont {T.}~\bibnamefont {Dutta}}, \bibinfo {author} {\bibfnamefont {P.}~\bibnamefont {H{\"a}nggi}}, \bibinfo {author} {\bibfnamefont {J.}~\bibnamefont {Gong}}, \bibinfo {author} {\bibfnamefont {D.}~\bibnamefont {Poletti}}, \ and\ \bibinfo {author} {\bibfnamefont {M.}~\bibnamefont {Mukherjee}},\ }\bibfield  {title} {\enquote {\bibinfo {title} {Single-atom energy-conversion device with a quantum load},}\ }\href@noop {} {\bibfield  {journal} {\bibinfo  {journal} {npj Quantum Information}\ }\textbf {\bibinfo {volume} {6}},\ \bibinfo {pages} {37} (\bibinfo {year} {2020})}\BibitemShut {NoStop}%
\bibitem [{\citenamefont {Boller}, \citenamefont {Imamo{\u{g}}lu},\ and\ \citenamefont {Harris}(1991)}]{boller1991observation}%
  \BibitemOpen
  \bibfield  {author} {\bibinfo {author} {\bibfnamefont {K.-J.}\ \bibnamefont {Boller}}, \bibinfo {author} {\bibfnamefont {A.}~\bibnamefont {Imamo{\u{g}}lu}}, \ and\ \bibinfo {author} {\bibfnamefont {S.~E.}\ \bibnamefont {Harris}},\ }\bibfield  {title} {\enquote {\bibinfo {title} {Observation of electromagnetically induced transparency},}\ }\href@noop {} {\bibfield  {journal} {\bibinfo  {journal} {Physical Review Letters}\ }\textbf {\bibinfo {volume} {66}},\ \bibinfo {pages} {2593} (\bibinfo {year} {1991})}\BibitemShut {NoStop}%
\bibitem [{\citenamefont {Harris}, \citenamefont {Field},\ and\ \citenamefont {Imamo{\u{g}}lu}(1990)}]{harris1990nonlinear}%
  \BibitemOpen
  \bibfield  {author} {\bibinfo {author} {\bibfnamefont {S.~E.}\ \bibnamefont {Harris}}, \bibinfo {author} {\bibfnamefont {J.}~\bibnamefont {Field}}, \ and\ \bibinfo {author} {\bibfnamefont {A.}~\bibnamefont {Imamo{\u{g}}lu}},\ }\bibfield  {title} {\enquote {\bibinfo {title} {Nonlinear optical processes using electromagnetically induced transparency},}\ }\href@noop {} {\bibfield  {journal} {\bibinfo  {journal} {Physical Review Letters}\ }\textbf {\bibinfo {volume} {64}},\ \bibinfo {pages} {1107} (\bibinfo {year} {1990})}\BibitemShut {NoStop}%
\bibitem [{\citenamefont {Harris}(2016{\natexlab{a}})}]{PhysRevA.94.053859_seharris}%
  \BibitemOpen
  \bibfield  {author} {\bibinfo {author} {\bibfnamefont {S.~E.}\ \bibnamefont {Harris}},\ }\bibfield  {title} {\enquote {\bibinfo {title} {Electromagnetically induced transparency and quantum heat engines},}\ }\href {\doibase 10.1103/PhysRevA.94.053859} {\bibfield  {journal} {\bibinfo  {journal} {Phys. Rev. A}\ }\textbf {\bibinfo {volume} {94}},\ \bibinfo {pages} {053859} (\bibinfo {year} {2016}{\natexlab{a}})}\BibitemShut {NoStop}%
\bibitem [{\citenamefont {Ma}\ \emph {et~al.}(2024)\citenamefont {Ma}, \citenamefont {Shen}, \citenamefont {Zhang},\ and\ \citenamefont {Wu}}]{PhysRevA.109.012207_Zhang}%
  \BibitemOpen
  \bibfield  {author} {\bibinfo {author} {\bibfnamefont {J.-Y.}\ \bibnamefont {Ma}}, \bibinfo {author} {\bibfnamefont {H.-Z.}\ \bibnamefont {Shen}}, \bibinfo {author} {\bibfnamefont {X.-J.}\ \bibnamefont {Zhang}}, \ and\ \bibinfo {author} {\bibfnamefont {J.-H.}\ \bibnamefont {Wu}},\ }\bibfield  {title} {\enquote {\bibinfo {title} {Single-atom quantum heat engine based on electromagnetically induced transparency},}\ }\href {\doibase 10.1103/PhysRevA.109.012207} {\bibfield  {journal} {\bibinfo  {journal} {Phys. Rev. A}\ }\textbf {\bibinfo {volume} {109}},\ \bibinfo {pages} {012207} (\bibinfo {year} {2024})}\BibitemShut {NoStop}%
\bibitem [{\citenamefont {Harris}(2016{\natexlab{b}})}]{PhysRevA.94.053859}%
  \BibitemOpen
  \bibfield  {author} {\bibinfo {author} {\bibfnamefont {S.~E.}\ \bibnamefont {Harris}},\ }\bibfield  {title} {\enquote {\bibinfo {title} {Electromagnetically induced transparency and quantum heat engines},}\ }\href {\doibase 10.1103/PhysRevA.94.053859} {\bibfield  {journal} {\bibinfo  {journal} {Phys. Rev. A}\ }\textbf {\bibinfo {volume} {94}},\ \bibinfo {pages} {053859} (\bibinfo {year} {2016}{\natexlab{b}})}\BibitemShut {NoStop}%
\bibitem [{\citenamefont {Laskar}(2024)}]{Laskar_2024}%
  \BibitemOpen
  \bibfield  {author} {\bibinfo {author} {\bibfnamefont {R.}~\bibnamefont {Laskar}},\ }\bibfield  {title} {\enquote {\bibinfo {title} {Proposal for composite quantum electromagnetically induced transparency heat engine coupled by a nanomechanical mirror},}\ }\href {\doibase 10.1088/1361-6455/ad2183} {\bibfield  {journal} {\bibinfo  {journal} {Journal of Physics B: Atomic, Molecular and Optical Physics}\ }\textbf {\bibinfo {volume} {57}},\ \bibinfo {pages} {025402} (\bibinfo {year} {2024})}\BibitemShut {NoStop}%
\bibitem [{\citenamefont {Niaz}\ \emph {et~al.}(2024)\citenamefont {Niaz}, \citenamefont {Chuang}, \citenamefont {Badshah},\ and\ \citenamefont {Rahmatullah}}]{NIAZ2024171816}%
  \BibitemOpen
  \bibfield  {author} {\bibinfo {author} {\bibfnamefont {L.}~\bibnamefont {Niaz}}, \bibinfo {author} {\bibfnamefont {Y.-L.}\ \bibnamefont {Chuang}}, \bibinfo {author} {\bibfnamefont {F.}~\bibnamefont {Badshah}}, \ and\ \bibinfo {author} {\bibnamefont {Rahmatullah}},\ }\bibfield  {title} {\enquote {\bibinfo {title} {Gain-assisted quantum heat engine based on electromagnetically induced transparency},}\ }\href {\doibase https://doi.org/10.1016/j.ijleo.2024.171816} {\bibfield  {journal} {\bibinfo  {journal} {Optik}\ }\textbf {\bibinfo {volume} {306}},\ \bibinfo {pages} {171816} (\bibinfo {year} {2024})}\BibitemShut {NoStop}%
\bibitem [{\citenamefont {Zou}\ \emph {et~al.}(2017)\citenamefont {Zou}, \citenamefont {Jiang}, \citenamefont {Mei}, \citenamefont {Guo},\ and\ \citenamefont {Du}}]{PhysRevLett.119.050602_zou}%
  \BibitemOpen
  \bibfield  {author} {\bibinfo {author} {\bibfnamefont {Y.}~\bibnamefont {Zou}}, \bibinfo {author} {\bibfnamefont {Y.}~\bibnamefont {Jiang}}, \bibinfo {author} {\bibfnamefont {Y.}~\bibnamefont {Mei}}, \bibinfo {author} {\bibfnamefont {X.}~\bibnamefont {Guo}}, \ and\ \bibinfo {author} {\bibfnamefont {S.}~\bibnamefont {Du}},\ }\bibfield  {title} {\enquote {\bibinfo {title} {Quantum heat engine using electromagnetically induced transparency},}\ }\href {\doibase 10.1103/PhysRevLett.119.050602} {\bibfield  {journal} {\bibinfo  {journal} {Phys. Rev. Lett.}\ }\textbf {\bibinfo {volume} {119}},\ \bibinfo {pages} {050602} (\bibinfo {year} {2017})}\BibitemShut {NoStop}%
\bibitem [{\citenamefont {Harris}, \citenamefont {Field},\ and\ \citenamefont {Imamo\ifmmode~\breve{g}\else \u{g}\fi{}lu}(1990)}]{PhysRevLett.64.1107_Imamo_2}%
  \BibitemOpen
  \bibfield  {author} {\bibinfo {author} {\bibfnamefont {S.~E.}\ \bibnamefont {Harris}}, \bibinfo {author} {\bibfnamefont {J.~E.}\ \bibnamefont {Field}}, \ and\ \bibinfo {author} {\bibfnamefont {A.}~\bibnamefont {Imamo\ifmmode~\breve{g}\else \u{g}\fi{}lu}},\ }\bibfield  {title} {\enquote {\bibinfo {title} {Nonlinear optical processes using electromagnetically induced transparency},}\ }\href {\doibase 10.1103/PhysRevLett.64.1107} {\bibfield  {journal} {\bibinfo  {journal} {Phys. Rev. Lett.}\ }\textbf {\bibinfo {volume} {64}},\ \bibinfo {pages} {1107--1110} (\bibinfo {year} {1990})}\BibitemShut {NoStop}%
\bibitem [{\citenamefont {Fleischhauer}, \citenamefont {Imamoglu},\ and\ \citenamefont {Marangos}(2005)}]{RevModPhys.77.633_Imamogl_1}%
  \BibitemOpen
  \bibfield  {author} {\bibinfo {author} {\bibfnamefont {M.}~\bibnamefont {Fleischhauer}}, \bibinfo {author} {\bibfnamefont {A.}~\bibnamefont {Imamoglu}}, \ and\ \bibinfo {author} {\bibfnamefont {J.~P.}\ \bibnamefont {Marangos}},\ }\bibfield  {title} {\enquote {\bibinfo {title} {Electromagnetically induced transparency: Optics in coherent media},}\ }\href {\doibase 10.1103/RevModPhys.77.633} {\bibfield  {journal} {\bibinfo  {journal} {Rev. Mod. Phys.}\ }\textbf {\bibinfo {volume} {77}},\ \bibinfo {pages} {633--673} (\bibinfo {year} {2005})}\BibitemShut {NoStop}%
\bibitem [{\citenamefont {Imamolu}, \citenamefont {Field},\ and\ \citenamefont {Harris}(1991)}]{PhysRevLett.66.1154_Imamo_3}%
  \BibitemOpen
  \bibfield  {author} {\bibinfo {author} {\bibfnamefont {A.}~\bibnamefont {Imamolu}}, \bibinfo {author} {\bibfnamefont {J.~E.}\ \bibnamefont {Field}}, \ and\ \bibinfo {author} {\bibfnamefont {S.~E.}\ \bibnamefont {Harris}},\ }\bibfield  {title} {\enquote {\bibinfo {title} {Lasers without inversion: A closed lifetime broadened system},}\ }\href {\doibase 10.1103/PhysRevLett.66.1154} {\bibfield  {journal} {\bibinfo  {journal} {Phys. Rev. Lett.}\ }\textbf {\bibinfo {volume} {66}},\ \bibinfo {pages} {1154--1156} (\bibinfo {year} {1991})}\BibitemShut {NoStop}%
\bibitem [{\citenamefont {Geusic}, \citenamefont {Schulz-DuBios},\ and\ \citenamefont {Scovil}(1967)}]{PhysRev.156.343_Geusic}%
  \BibitemOpen
  \bibfield  {author} {\bibinfo {author} {\bibfnamefont {J.~E.}\ \bibnamefont {Geusic}}, \bibinfo {author} {\bibfnamefont {E.~O.}\ \bibnamefont {Schulz-DuBios}}, \ and\ \bibinfo {author} {\bibfnamefont {H.~E.~D.}\ \bibnamefont {Scovil}},\ }\bibfield  {title} {\enquote {\bibinfo {title} {Quantum equivalent of the carnot cycle},}\ }\href {\doibase 10.1103/PhysRev.156.343} {\bibfield  {journal} {\bibinfo  {journal} {Phys. Rev.}\ }\textbf {\bibinfo {volume} {156}},\ \bibinfo {pages} {343--351} (\bibinfo {year} {1967})}\BibitemShut {NoStop}%
\bibitem [{\citenamefont {Mihalas}\ and\ \citenamefont {Mihalas}(1984)}]{0-19-503437-6_b_w_deri_book}%
  \BibitemOpen
  \bibfield  {author} {\bibinfo {author} {\bibfnamefont {D.}~\bibnamefont {Mihalas}}\ and\ \bibinfo {author} {\bibfnamefont {B.~W.}\ \bibnamefont {Mihalas}},\ }\href@noop {} {\emph {\bibinfo {title} {Foundations of Radiation Hydrodynamics}}}\ (\bibinfo  {publisher} {Oxford University Press, New York},\ \bibinfo {year} {1984})\ pp.\ \bibinfo {pages} {xv+718}\BibitemShut {NoStop}%
\bibitem [{\citenamefont {{\c{C}}akmak}(2020)}]{ccakmak2020ergotropy_coh}%
  \BibitemOpen
  \bibfield  {author} {\bibinfo {author} {\bibfnamefont {B.}~\bibnamefont {{\c{C}}akmak}},\ }\bibfield  {title} {\enquote {\bibinfo {title} {Ergotropy from coherences in an open quantum system},}\ }\href@noop {} {\bibfield  {journal} {\bibinfo  {journal} {Physical Review E}\ }\textbf {\bibinfo {volume} {102}},\ \bibinfo {pages} {042111} (\bibinfo {year} {2020})}\BibitemShut {NoStop}%
\bibitem [{\citenamefont {Biswas}\ \emph {et~al.}(2022)\citenamefont {Biswas}, \citenamefont {{\L}obejko}, \citenamefont {Mazurek}, \citenamefont {Ja{\l}owiecki},\ and\ \citenamefont {Horodecki}}]{biswas2022extraction_erg}%
  \BibitemOpen
  \bibfield  {author} {\bibinfo {author} {\bibfnamefont {T.}~\bibnamefont {Biswas}}, \bibinfo {author} {\bibfnamefont {M.}~\bibnamefont {{\L}obejko}}, \bibinfo {author} {\bibfnamefont {P.}~\bibnamefont {Mazurek}}, \bibinfo {author} {\bibfnamefont {K.}~\bibnamefont {Ja{\l}owiecki}}, \ and\ \bibinfo {author} {\bibfnamefont {M.}~\bibnamefont {Horodecki}},\ }\bibfield  {title} {\enquote {\bibinfo {title} {Extraction of ergotropy: Free energy bound and application to open cycle engines},}\ }\href@noop {} {\bibfield  {journal} {\bibinfo  {journal} {Quantum}\ }\textbf {\bibinfo {volume} {6}},\ \bibinfo {pages} {841} (\bibinfo {year} {2022})}\BibitemShut {NoStop}%
\bibitem [{\citenamefont {Sarmah}\ and\ \citenamefont {Goswami}(2024{\natexlab{c}})}]{PhysRevA.110.032213_mjs_erg}%
  \BibitemOpen
  \bibfield  {author} {\bibinfo {author} {\bibfnamefont {M.~J.}\ \bibnamefont {Sarmah}}\ and\ \bibinfo {author} {\bibfnamefont {H.~P.}\ \bibnamefont {Goswami}},\ }\bibfield  {title} {\enquote {\bibinfo {title} {Noise-induced coherent ergotropy of a quantum heat engine},}\ }\href {\doibase 10.1103/PhysRevA.110.032213} {\bibfield  {journal} {\bibinfo  {journal} {Phys. Rev. A}\ }\textbf {\bibinfo {volume} {110}},\ \bibinfo {pages} {032213} (\bibinfo {year} {2024}{\natexlab{c}})}\BibitemShut {NoStop}%
\bibitem [{\citenamefont {Elouard}\ and\ \citenamefont {Lombard~Latune}(2023)}]{elouard2023extending_exp_erg}%
  \BibitemOpen
  \bibfield  {author} {\bibinfo {author} {\bibfnamefont {C.}~\bibnamefont {Elouard}}\ and\ \bibinfo {author} {\bibfnamefont {C.}~\bibnamefont {Lombard~Latune}},\ }\bibfield  {title} {\enquote {\bibinfo {title} {Extending the laws of thermodynamics for arbitrary autonomous quantum systems},}\ }\href@noop {} {\bibfield  {journal} {\bibinfo  {journal} {PRX Quantum}\ }\textbf {\bibinfo {volume} {4}},\ \bibinfo {pages} {020309} (\bibinfo {year} {2023})}\BibitemShut {NoStop}%
\bibitem [{vsi(2017)}]{vsibalic2017arc_ARC}%
  \BibitemOpen
  \bibfield  {title} {\enquote {\bibinfo {title} {Arc: An open-source library for calculating properties of alkali rydberg atoms},}\ }\href@noop {} {\bibfield  {journal} {\bibinfo  {journal} {Computer Physics Communications}\ }\textbf {\bibinfo {volume} {220}},\ \bibinfo {pages} {319--331} (\bibinfo {year} {2017})}\BibitemShut {NoStop}%
\bibitem [{\citenamefont {Lai}\ \emph {et~al.}(2018)\citenamefont {Lai}, \citenamefont {Zhang}, \citenamefont {Gou},\ and\ \citenamefont {Li}}]{PhysRevA.98.052503}%
  \BibitemOpen
  \bibfield  {author} {\bibinfo {author} {\bibfnamefont {Z.}~\bibnamefont {Lai}}, \bibinfo {author} {\bibfnamefont {S.}~\bibnamefont {Zhang}}, \bibinfo {author} {\bibfnamefont {Q.}~\bibnamefont {Gou}}, \ and\ \bibinfo {author} {\bibfnamefont {Y.}~\bibnamefont {Li}},\ }\bibfield  {title} {\enquote {\bibinfo {title} {Polarizabilities of rydberg states of rb atoms with $n$ up to 140},}\ }\href {\doibase 10.1103/PhysRevA.98.052503} {\bibfield  {journal} {\bibinfo  {journal} {Phys. Rev. A}\ }\textbf {\bibinfo {volume} {98}},\ \bibinfo {pages} {052503} (\bibinfo {year} {2018})}\BibitemShut {NoStop}%
\bibitem [{\citenamefont {Krug}(2001)}]{MPQ-274}%
  \BibitemOpen
  \bibfield  {author} {\bibinfo {author} {\bibfnamefont {A.}~\bibnamefont {Krug}},\ }\href {http://nbn-resolving.de/urn:nbn:de:bvb:19-3362} {\enquote {\bibinfo {title} {Alkali rydberg states in electromagnetic fields},}\ } (\bibinfo {year} {2001})\BibitemShut {NoStop}%
\bibitem [{\citenamefont {Theodosiou}(1984{\natexlab{a}})}]{PhysRevA.30.2881}%
  \BibitemOpen
  \bibfield  {author} {\bibinfo {author} {\bibfnamefont {C.~E.}\ \bibnamefont {Theodosiou}},\ }\bibfield  {title} {\enquote {\bibinfo {title} {Lifetimes of alkali-metal---atom rydberg states},}\ }\href {\doibase 10.1103/PhysRevA.30.2881} {\bibfield  {journal} {\bibinfo  {journal} {Phys. Rev. A}\ }\textbf {\bibinfo {volume} {30}},\ \bibinfo {pages} {2881--2909} (\bibinfo {year} {1984}{\natexlab{a}})}\BibitemShut {NoStop}%
\bibitem [{\citenamefont {van Raan}, \citenamefont {Baum},\ and\ \citenamefont {Raith}(1976)}]{JPhysB1976}%
  \BibitemOpen
  \bibfield  {author} {\bibinfo {author} {\bibfnamefont {A.~F.~J.}\ \bibnamefont {van Raan}}, \bibinfo {author} {\bibfnamefont {G.}~\bibnamefont {Baum}}, \ and\ \bibinfo {author} {\bibfnamefont {W.}~\bibnamefont {Raith}},\ }\bibfield  {title} {\enquote {\bibinfo {title} {Measurement of the scalar polarizability of very highly excited states of caesium},}\ }\href {\doibase 10.1088/0022-3700/9/12/003} {\bibfield  {journal} {\bibinfo  {journal} {Journal of Physics B: Atomic and Molecular Physics}\ }\textbf {\bibinfo {volume} {9}},\ \bibinfo {pages} {L349} (\bibinfo {year} {1976})}\BibitemShut {NoStop}%
\bibitem [{\citenamefont {Jones}, \citenamefont {Potvliege},\ and\ \citenamefont {Spannowsky}(2020{\natexlab{a}})}]{PhysRevResearch.2.013244_H_atom_exp}%
  \BibitemOpen
  \bibfield  {author} {\bibinfo {author} {\bibfnamefont {M.~P.~A.}\ \bibnamefont {Jones}}, \bibinfo {author} {\bibfnamefont {R.~M.}\ \bibnamefont {Potvliege}}, \ and\ \bibinfo {author} {\bibfnamefont {M.}~\bibnamefont {Spannowsky}},\ }\bibfield  {title} {\enquote {\bibinfo {title} {Probing new physics using rydberg states of atomic hydrogen},}\ }\href {\doibase 10.1103/PhysRevResearch.2.013244} {\bibfield  {journal} {\bibinfo  {journal} {Phys. Rev. Res.}\ }\textbf {\bibinfo {volume} {2}},\ \bibinfo {pages} {013244} (\bibinfo {year} {2020}{\natexlab{a}})}\BibitemShut {NoStop}%
\bibitem [{\citenamefont {Theodosiou}(1984{\natexlab{b}})}]{PhysRevA.30.2881_all_alklai_exp_access}%
  \BibitemOpen
  \bibfield  {author} {\bibinfo {author} {\bibfnamefont {C.~E.}\ \bibnamefont {Theodosiou}},\ }\bibfield  {title} {\enquote {\bibinfo {title} {Lifetimes of alkali-metal---atom rydberg states},}\ }\href {\doibase 10.1103/PhysRevA.30.2881} {\bibfield  {journal} {\bibinfo  {journal} {Phys. Rev. A}\ }\textbf {\bibinfo {volume} {30}},\ \bibinfo {pages} {2881--2909} (\bibinfo {year} {1984}{\natexlab{b}})}\BibitemShut {NoStop}%
\bibitem [{\citenamefont {Zhao}\ \emph {et~al.}(2009)\citenamefont {Zhao}, \citenamefont {Zhu}, \citenamefont {Zhang}, \citenamefont {Feng}, \citenamefont {Li},\ and\ \citenamefont {Jia}}]{OE.17.015821}%
  \BibitemOpen
  \bibfield  {author} {\bibinfo {author} {\bibfnamefont {J.}~\bibnamefont {Zhao}}, \bibinfo {author} {\bibfnamefont {X.}~\bibnamefont {Zhu}}, \bibinfo {author} {\bibfnamefont {L.}~\bibnamefont {Zhang}}, \bibinfo {author} {\bibfnamefont {Z.}~\bibnamefont {Feng}}, \bibinfo {author} {\bibfnamefont {C.}~\bibnamefont {Li}}, \ and\ \bibinfo {author} {\bibfnamefont {S.}~\bibnamefont {Jia}},\ }\bibfield  {title} {\enquote {\bibinfo {title} {High sensitivity spectroscopy of cesium rydberg atoms using electromagnetically induced transparency},}\ }\href {\doibase 10.1364/OE.17.015821} {\bibfield  {journal} {\bibinfo  {journal} {Opt. Express}\ }\textbf {\bibinfo {volume} {17}},\ \bibinfo {pages} {15821--15826} (\bibinfo {year} {2009})}\BibitemShut {NoStop}%
\bibitem [{\citenamefont {Jones}, \citenamefont {Potvliege},\ and\ \citenamefont {Spannowsky}(2020{\natexlab{b}})}]{PhysRevResearch.2.013244}%
  \BibitemOpen
  \bibfield  {author} {\bibinfo {author} {\bibfnamefont {M.~P.~A.}\ \bibnamefont {Jones}}, \bibinfo {author} {\bibfnamefont {R.~M.}\ \bibnamefont {Potvliege}}, \ and\ \bibinfo {author} {\bibfnamefont {M.}~\bibnamefont {Spannowsky}},\ }\bibfield  {title} {\enquote {\bibinfo {title} {Probing new physics using rydberg states of atomic hydrogen},}\ }\href {\doibase 10.1103/PhysRevResearch.2.013244} {\bibfield  {journal} {\bibinfo  {journal} {Phys. Rev. Res.}\ }\textbf {\bibinfo {volume} {2}},\ \bibinfo {pages} {013244} (\bibinfo {year} {2020}{\natexlab{b}})}\BibitemShut {NoStop}%
\bibitem [{\citenamefont {Micha{\'n}k{\'o}w}, \citenamefont {Sakowski},\ and\ \citenamefont {{\'S}lepaczuk}(2024)}]{michankow2024mean_loss_mae}%
  \BibitemOpen
  \bibfield  {author} {\bibinfo {author} {\bibfnamefont {J.}~\bibnamefont {Micha{\'n}k{\'o}w}}, \bibinfo {author} {\bibfnamefont {P.}~\bibnamefont {Sakowski}}, \ and\ \bibinfo {author} {\bibfnamefont {R.}~\bibnamefont {{\'S}lepaczuk}},\ }\bibfield  {title} {\enquote {\bibinfo {title} {Mean absolute directional loss as a new loss function for machine learning problems in algorithmic investment strategies},}\ }\href@noop {} {\bibfield  {journal} {\bibinfo  {journal} {Journal of Computational Science}\ ,\ \bibinfo {pages} {102375}} (\bibinfo {year} {2024})}\BibitemShut {NoStop}%
\bibitem [{\citenamefont {Qi}\ \emph {et~al.}(2020)\citenamefont {Qi}, \citenamefont {Du}, \citenamefont {Siniscalchi}, \citenamefont {Ma},\ and\ \citenamefont {Lee}}]{qi2020mean_MAE}%
  \BibitemOpen
  \bibfield  {author} {\bibinfo {author} {\bibfnamefont {J.}~\bibnamefont {Qi}}, \bibinfo {author} {\bibfnamefont {J.}~\bibnamefont {Du}}, \bibinfo {author} {\bibfnamefont {S.~M.}\ \bibnamefont {Siniscalchi}}, \bibinfo {author} {\bibfnamefont {X.}~\bibnamefont {Ma}}, \ and\ \bibinfo {author} {\bibfnamefont {C.-H.}\ \bibnamefont {Lee}},\ }\bibfield  {title} {\enquote {\bibinfo {title} {On mean absolute error for deep neural network based vector-to-vector regression},}\ }\href@noop {} {\bibfield  {journal} {\bibinfo  {journal} {IEEE Signal Processing Letters}\ }\textbf {\bibinfo {volume} {27}},\ \bibinfo {pages} {1485--1489} (\bibinfo {year} {2020})}\BibitemShut {NoStop}%
\bibitem [{\citenamefont {Abadi}\ \emph {et~al.}(2015)\citenamefont {Abadi}, \citenamefont {Agarwal}, \citenamefont {Barham}, \citenamefont {Brevdo}, \citenamefont {Chen}, \citenamefont {Citro}, \citenamefont {Corrado}, \citenamefont {Davis}, \citenamefont {Dean}, \citenamefont {Devin}, \citenamefont {Ghemawat}, \citenamefont {Goodfellow}, \citenamefont {Harp}, \citenamefont {Irving}, \citenamefont {Isard}, \citenamefont {Jia}, \citenamefont {Jozefowicz}, \citenamefont {Kaiser}, \citenamefont {Kudlur}, \citenamefont {Levenberg}, \citenamefont {Man\'{e}}, \citenamefont {Monga}, \citenamefont {Moore}, \citenamefont {Murray}, \citenamefont {Olah}, \citenamefont {Schuster}, \citenamefont {Shlens}, \citenamefont {Steiner}, \citenamefont {Sutskever}, \citenamefont {Talwar}, \citenamefont {Tucker}, \citenamefont {Vanhoucke}, \citenamefont {Vasudevan}, \citenamefont {Vi\'{e}gas}, \citenamefont {Vinyals}, \citenamefont {Warden}, \citenamefont {Wattenberg}, \citenamefont {Wicke}, \citenamefont {Yu},\ and\ \citenamefont
  {Zheng}}]{tensorflow2015-whitepaper}%
  \BibitemOpen
  \bibfield  {author} {\bibinfo {author} {\bibfnamefont {M.}~\bibnamefont {Abadi}}, \bibinfo {author} {\bibfnamefont {A.}~\bibnamefont {Agarwal}}, \bibinfo {author} {\bibfnamefont {P.}~\bibnamefont {Barham}}, \bibinfo {author} {\bibfnamefont {E.}~\bibnamefont {Brevdo}}, \bibinfo {author} {\bibfnamefont {Z.}~\bibnamefont {Chen}}, \bibinfo {author} {\bibfnamefont {C.}~\bibnamefont {Citro}}, \bibinfo {author} {\bibfnamefont {G.~S.}\ \bibnamefont {Corrado}}, \bibinfo {author} {\bibfnamefont {A.}~\bibnamefont {Davis}}, \bibinfo {author} {\bibfnamefont {J.}~\bibnamefont {Dean}}, \bibinfo {author} {\bibfnamefont {M.}~\bibnamefont {Devin}}, \bibinfo {author} {\bibfnamefont {S.}~\bibnamefont {Ghemawat}}, \bibinfo {author} {\bibfnamefont {I.}~\bibnamefont {Goodfellow}}, \bibinfo {author} {\bibfnamefont {A.}~\bibnamefont {Harp}}, \bibinfo {author} {\bibfnamefont {G.}~\bibnamefont {Irving}}, \bibinfo {author} {\bibfnamefont {M.}~\bibnamefont {Isard}}, \bibinfo {author} {\bibfnamefont {Y.}~\bibnamefont {Jia}}, \bibinfo
  {author} {\bibfnamefont {R.}~\bibnamefont {Jozefowicz}}, \bibinfo {author} {\bibfnamefont {L.}~\bibnamefont {Kaiser}}, \bibinfo {author} {\bibfnamefont {M.}~\bibnamefont {Kudlur}}, \bibinfo {author} {\bibfnamefont {J.}~\bibnamefont {Levenberg}}, \bibinfo {author} {\bibfnamefont {D.}~\bibnamefont {Man\'{e}}}, \bibinfo {author} {\bibfnamefont {R.}~\bibnamefont {Monga}}, \bibinfo {author} {\bibfnamefont {S.}~\bibnamefont {Moore}}, \bibinfo {author} {\bibfnamefont {D.}~\bibnamefont {Murray}}, \bibinfo {author} {\bibfnamefont {C.}~\bibnamefont {Olah}}, \bibinfo {author} {\bibfnamefont {M.}~\bibnamefont {Schuster}}, \bibinfo {author} {\bibfnamefont {J.}~\bibnamefont {Shlens}}, \bibinfo {author} {\bibfnamefont {B.}~\bibnamefont {Steiner}}, \bibinfo {author} {\bibfnamefont {I.}~\bibnamefont {Sutskever}}, \bibinfo {author} {\bibfnamefont {K.}~\bibnamefont {Talwar}}, \bibinfo {author} {\bibfnamefont {P.}~\bibnamefont {Tucker}}, \bibinfo {author} {\bibfnamefont {V.}~\bibnamefont {Vanhoucke}}, \bibinfo {author}
  {\bibfnamefont {V.}~\bibnamefont {Vasudevan}}, \bibinfo {author} {\bibfnamefont {F.}~\bibnamefont {Vi\'{e}gas}}, \bibinfo {author} {\bibfnamefont {O.}~\bibnamefont {Vinyals}}, \bibinfo {author} {\bibfnamefont {P.}~\bibnamefont {Warden}}, \bibinfo {author} {\bibfnamefont {M.}~\bibnamefont {Wattenberg}}, \bibinfo {author} {\bibfnamefont {M.}~\bibnamefont {Wicke}}, \bibinfo {author} {\bibfnamefont {Y.}~\bibnamefont {Yu}}, \ and\ \bibinfo {author} {\bibfnamefont {X.}~\bibnamefont {Zheng}},\ }\href {https://www.tensorflow.org/} {\enquote {\bibinfo {title} {{TensorFlow}: Large-scale machine learning on heterogeneous systems},}\ } (\bibinfo {year} {2015}),\ \bibinfo {note} {software available from tensorflow.org}\BibitemShut {NoStop}%
\bibitem [{\citenamefont {Goodfellow}, \citenamefont {Bengio},\ and\ \citenamefont {Courville}(2016)}]{Goodfellow-et-al-2016_deep_learning}%
  \BibitemOpen
  \bibfield  {author} {\bibinfo {author} {\bibfnamefont {I.}~\bibnamefont {Goodfellow}}, \bibinfo {author} {\bibfnamefont {Y.}~\bibnamefont {Bengio}}, \ and\ \bibinfo {author} {\bibfnamefont {A.}~\bibnamefont {Courville}},\ }\href@noop {} {\emph {\bibinfo {title} {Deep Learning}}}\ (\bibinfo  {publisher} {MIT Press},\ \bibinfo {year} {2016})\ \bibinfo {note} {\url{http://www.deeplearningbook.org}}\BibitemShut {NoStop}%
\bibitem [{\citenamefont {Seyrek}\ and\ \citenamefont {Uysal}(2024)}]{Seyrek2024_act_fns)}%
  \BibitemOpen
  \bibfield  {author} {\bibinfo {author} {\bibfnamefont {E.~C.}\ \bibnamefont {Seyrek}}\ and\ \bibinfo {author} {\bibfnamefont {M.}~\bibnamefont {Uysal}},\ }\bibfield  {title} {\enquote {\bibinfo {title} {A comparative analysis of various activation functions and optimizers in a convolutional neural network for hyperspectral image classification},}\ }\href {\doibase 10.1007/s11042-023-17546-5} {\bibfield  {journal} {\bibinfo  {journal} {Multimedia Tools and Applications}\ }\textbf {\bibinfo {volume} {83}},\ \bibinfo {pages} {53785--53816} (\bibinfo {year} {2024})}\BibitemShut {NoStop}%
\bibitem [{\citenamefont {Sarmah}(2024)}]{githubGitHubManashSarmahDLEnabledPredictionofQHEsBasedonEIT}%
  \BibitemOpen
  \bibfield  {author} {\bibinfo {author} {\bibfnamefont {M.~J.}\ \bibnamefont {Sarmah}},\ }\href@noop {} {\enquote {\bibinfo {title} {{D}{L}-{E}nabled-{P}rediction-of-{Q}{H}{E}s-{B}ased-on-{E}{I}{T}: {D}eep {L}earning-{E}nabled {P}rediction of {Q}uantum {H}eat {E}ngines {B}ased on {E}lectromagnetically {I}nduced {T}ransparency --- github.com},}\ }\bibinfo {howpublished} {\url{https://github.com/ManashSarmah/DL-Enabled-Prediction-of-QHEs-Based-on-EIT.git}} (\bibinfo {year} {2024})\BibitemShut {NoStop}%
\bibitem [{\citenamefont {Mungan}(2005)}]{10.1119/1.1842732_Mungan_th}%
  \BibitemOpen
  \bibfield  {author} {\bibinfo {author} {\bibfnamefont {C.~E.}\ \bibnamefont {Mungan}},\ }\bibfield  {title} {\enquote {\bibinfo {title} {{Radiation thermodynamics with applications to lasing and fluorescent cooling}},}\ }\href {\doibase 10.1119/1.1842732} {\bibfield  {journal} {\bibinfo  {journal} {American Journal of Physics}\ }\textbf {\bibinfo {volume} {73}},\ \bibinfo {pages} {315--322} (\bibinfo {year} {2005})},\ \Eprint {http://arxiv.org/abs/https://pubs.aip.org/aapt/ajp/article-pdf/73/4/315/7521568/315\_1\_online.pdf} {https://pubs.aip.org/aapt/ajp/article-pdf/73/4/315/7521568/315\_1\_online.pdf} \BibitemShut {NoStop}%
\bibitem [{\citenamefont {Ruan}, \citenamefont {Rand},\ and\ \citenamefont {Kaviany}(2007)}]{PhysRevB.75.214304_qh_th_entropy}%
  \BibitemOpen
  \bibfield  {author} {\bibinfo {author} {\bibfnamefont {X.~L.}\ \bibnamefont {Ruan}}, \bibinfo {author} {\bibfnamefont {S.~C.}\ \bibnamefont {Rand}}, \ and\ \bibinfo {author} {\bibfnamefont {M.}~\bibnamefont {Kaviany}},\ }\bibfield  {title} {\enquote {\bibinfo {title} {Entropy and efficiency in laser cooling of solids},}\ }\href {\doibase 10.1103/PhysRevB.75.214304} {\bibfield  {journal} {\bibinfo  {journal} {Phys. Rev. B}\ }\textbf {\bibinfo {volume} {75}},\ \bibinfo {pages} {214304} (\bibinfo {year} {2007})}\BibitemShut {NoStop}%
\bibitem [{\citenamefont {Dubey}, \citenamefont {Singh},\ and\ \citenamefont {Chaudhuri}(2022)}]{dubey2022activation_tanh_activation_fncs}%
  \BibitemOpen
  \bibfield  {author} {\bibinfo {author} {\bibfnamefont {S.~R.}\ \bibnamefont {Dubey}}, \bibinfo {author} {\bibfnamefont {S.~K.}\ \bibnamefont {Singh}}, \ and\ \bibinfo {author} {\bibfnamefont {B.~B.}\ \bibnamefont {Chaudhuri}},\ }\bibfield  {title} {\enquote {\bibinfo {title} {Activation functions in deep learning: A comprehensive survey and benchmark},}\ }\href@noop {} {\bibfield  {journal} {\bibinfo  {journal} {Neurocomputing}\ }\textbf {\bibinfo {volume} {503}},\ \bibinfo {pages} {92--108} (\bibinfo {year} {2022})}\BibitemShut {NoStop}%
\bibitem [{\citenamefont {Serengil}(2017)}]{sefiks5597_tanh}%
  \BibitemOpen
  \bibfield  {author} {\bibinfo {author} {\bibfnamefont {S.~I.}\ \bibnamefont {Serengil}},\ }\href@noop {} {\enquote {\bibinfo {title} {Hyperbolic tangent as neural network activation function},}\ }\bibinfo {howpublished} {https://sefiks.com/2017/01/29/hyperbolic-tangent-as-neural-network-activation-function/} (\bibinfo {year} {2017}),\ \bibinfo {note} {[Online; accessed 2024-09-24]}\BibitemShut {NoStop}%
\bibitem [{\citenamefont {Szanda{\l}a}(2021)}]{Szandała2021_act_fns}%
  \BibitemOpen
  \bibfield  {author} {\bibinfo {author} {\bibfnamefont {T.}~\bibnamefont {Szanda{\l}a}},\ }\enquote {\bibinfo {title} {Review and comparison of commonly used activation functions for deep neural networks},}\ in\ \href {\doibase 10.1007/978-981-15-5495-7_11} {\emph {\bibinfo {booktitle} {Bio-inspired Neurocomputing}}},\ \bibinfo {editor} {edited by\ \bibinfo {editor} {\bibfnamefont {A.~K.}\ \bibnamefont {Bhoi}}, \bibinfo {editor} {\bibfnamefont {P.~K.}\ \bibnamefont {Mallick}}, \bibinfo {editor} {\bibfnamefont {C.-M.}\ \bibnamefont {Liu}}, \ and\ \bibinfo {editor} {\bibfnamefont {V.~E.}\ \bibnamefont {Balas}}}\ (\bibinfo  {publisher} {Springer Singapore},\ \bibinfo {address} {Singapore},\ \bibinfo {year} {2021})\ pp.\ \bibinfo {pages} {203--224}\BibitemShut {NoStop}%
\bibitem [{\citenamefont {Maurya}\ and\ \citenamefont {Yadav}(2023)}]{10.1007/978-981-99-3432-4_7_adam_optimiser}%
  \BibitemOpen
  \bibfield  {author} {\bibinfo {author} {\bibfnamefont {M.}~\bibnamefont {Maurya}}\ and\ \bibinfo {author} {\bibfnamefont {N.}~\bibnamefont {Yadav}},\ }\bibfield  {title} {\enquote {\bibinfo {title} {A comparative analysis of gradient-based optimization methods for machine learning problems},}\ }in\ \href@noop {} {\emph {\bibinfo {booktitle} {Proceedings on International Conference on Data Analytics and Computing}}},\ \bibinfo {editor} {edited by\ \bibinfo {editor} {\bibfnamefont {A.}~\bibnamefont {Yadav}}, \bibinfo {editor} {\bibfnamefont {G.}~\bibnamefont {Gupta}}, \bibinfo {editor} {\bibfnamefont {P.}~\bibnamefont {Rana}}, \ and\ \bibinfo {editor} {\bibfnamefont {J.~H.}\ \bibnamefont {Kim}}}\ (\bibinfo  {publisher} {Springer Nature Singapore},\ \bibinfo {address} {Singapore},\ \bibinfo {year} {2023})\ pp.\ \bibinfo {pages} {85--102}\BibitemShut {NoStop}%
\end{thebibliography}%


%

\end{document}